\DocumentMetadata{}

\documentclass[sigconf]{acmart}

\usepackage{bm} 
\usepackage{boxedminipage}
\usepackage{color}
\usepackage{colortbl}
\usepackage{float}
\usepackage{lscape}
\usepackage{makecell}
\usepackage{multirow}
\usepackage{subcaption}
\usepackage{soul}

\newcommand{\proposed}{IMUCoCo}

\newcommand{\tabref}[1]{Table~\ref{#1}}
\newcommand{\figref}[1]{Figure~\ref{#1}}
\newcommand{\secref}[1]{Section~\ref{#1}}

\newcommand{\appref}[1]{Appendix~\ref{#1}}

\newcolumntype{C}[1]{>{\centering\arraybackslash}m{#1}}

\newcommand{\etal}{\textit{et al.}}
\newcommand{\eg}{\textit{e.g.},~}
\newcommand{\ie}{\textit{i.e.},~}

\usepackage{amsmath}
\DeclareMathOperator*{\argmin}{argmin}

\AtBeginDocument{%
  \providecommand\BibTeX{{%
    \normalfont B\kern-0.5em{\scshape i\kern-0.25em b}\kern-0.8em\TeX}}}

\copyrightyear{2025}
\acmYear{2025}
\setcopyright{cc}
\setcctype{by}
\acmConference[UIST '25]{The 38th Annual ACM Symposium on User Interface Software and Technology}{September 28-October 1, 2025}{Busan, Republic of Korea}
\acmBooktitle{The 38th Annual ACM Symposium on User Interface Software and Technology (UIST '25), September 28-October 1, 2025, Busan, Republic of Korea}\acmDOI{10.1145/3746059.3747695}
\acmISBN{979-8-4007-2037-6/2025/09}

\begin{document}

\title[\proposed{}: Enabling Flexible On-Body IMU Placement ...]{\proposed{}: Enabling Flexible On-Body IMU Placement\\for Human Pose Estimation and Activity Recognition}


\author{Haozhe Zhou}
\orcid{0009-0002-7315-6033}
\affiliation{
  \institution{Carnegie Mellon University}
  \city{Pittsburgh}
  \state{PA}
  \country{USA}
}
\email{haozhezh@cs.cmu.edu}

\author{Riku Arakawa}
\orcid{0000-0001-7868-4754}
\affiliation{
  \institution{Carnegie Mellon University}
  \city{Pittsburgh}
  \state{PA}
  \country{USA}
}
\email{rarakawa@cs.cmu.edu}

\author{Yuvraj Agarwal}
\orcid{0000-0001-9304-6080}
\affiliation{
  \institution{Carnegie Mellon University}
  \city{Pittsburgh}
  \state{PA}
  \country{USA}
}
\email{yuvraj@cs.cmu.edu}

\author{Mayank Goel}
\orcid{0000-0003-1237-7545}
\affiliation{
  \institution{Carnegie Mellon University}
  \city{Pittsburgh}
  \state{PA}
  \country{USA}
}
\email{mayankgoel@cmu.edu}

\renewcommand{\shortauthors}{Zhou et al.}

\newcommand{\ProposedOurDatasetStandardGAE}{27.6}
\newcommand{\MobilePoserOurDatasetStandardGAE}{31.8}

\begin{teaserfigure}
    \centering
    \includegraphics[width=0.90\linewidth]{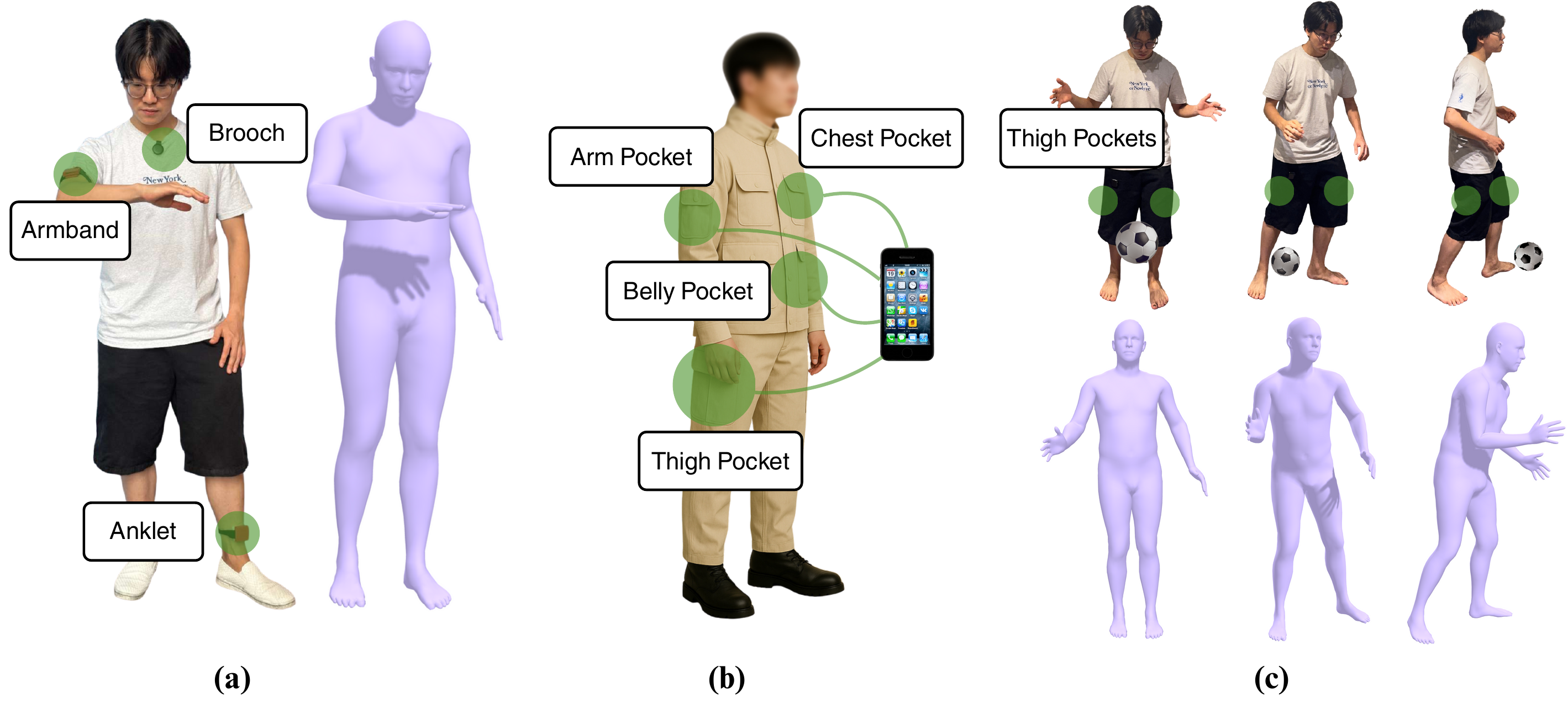}
    \caption{(a) ~\proposed{} enables pose estimation from atypical locations like the upper arm, upper chest, and ankle.~ (b) ~\proposed{} allows users to put their IMU sensing devices in different pockets of their clothing for convenience. (c) With \proposed{}, users can move the sensors to appropriate locations according to different application requirements (\eg IMUs in the thigh pockets to track leg movements during soccer).}
    \Description{Illustration of wearable sensor and smartphone placement variations across different conditions. (A) Shows three wearable sensor placements on a person: Brooch (upper chest), Armband (upper arm), and Anklet (lower leg). These locations are annotated with green circles and corresponding labels. (B) Depicts a person in workwear, indicating four possible smartphone locations: Arm Pocket, Chest Pocket, Belly Pocket, and Thigh Pocket. The smartphone is shown floating and linked with green lines to each pocket position. (C) Demonstrates a dynamic movement (\eg playing with a ball) with three body poses and corresponding sensor placements marked on the thighs of a purple human model below each pose. This suggests tracking during physical activities.}
    \label{fig:teaser}
\end{teaserfigure}
\begin{abstract} 
IMUs are regularly used to sense human motion, recognize activities, and estimate full-body pose. 
Users are typically required to place sensors in predefined locations that are often dictated by common wearable form factors and the machine learning model's training process. 
Consequently, despite the increasing number of everyday devices equipped with IMUs, the limited adaptability has significantly constrained the user experience to only using a few well-explored device placements (\eg wrist and ears).
In this paper, we rethink IMU-based motion sensing by acknowledging that signals can be captured from any point on the human body. 
We introduce \textbf{IMU} over \textbf{Co}ntinuous \textbf{Co}ordinates (\proposed{}), a novel framework that maps signals from a variable number of IMUs placed on the body surface into a unified feature space based on their spatial coordinates. 
These features can be plugged into downstream models for pose estimation and activity recognition.
Our evaluations demonstrate that \proposed{} supports accurate pose estimation in a wide range of typical and atypical sensor placements.
Overall, \proposed{} supports significantly more flexible use of IMUs for motion sensing than the state-of-the-art, allowing users to place their sensors-laden devices according to their needs and preferences. 
The framework also supports the ability to change device locations depending on the context and suggests placement depending on the use case.
\end{abstract}



\maketitle

\section{Introduction}
\label{sec:intro}

The ubiquity offered by IMUs has made them an attractive sensor for sensing human motion and pose~\cite{DBLP:conf/chi/AhujaMG021,DBLP:conf/eccv/JiangSQFLSH22}. 
Consumer IMUs are used for many applications, such as gait analysis using foot-mounted devices~\cite{Yamamoto2022Verification}, fall risk estimation using a smartphone~\cite{Isho2015}, hyperactivity monitoring using smartwatches~\cite{DBLP:journals/imwut/ArakawaAMTSLG23}, and human pose estimation using different consumer devices~\cite{DBLP:conf/uist/XuGHA24,DBLP:conf/cvpr/WouweLFDL24}. 

\begin{figure*}[t]
    \centering
    \includegraphics[width=\linewidth]{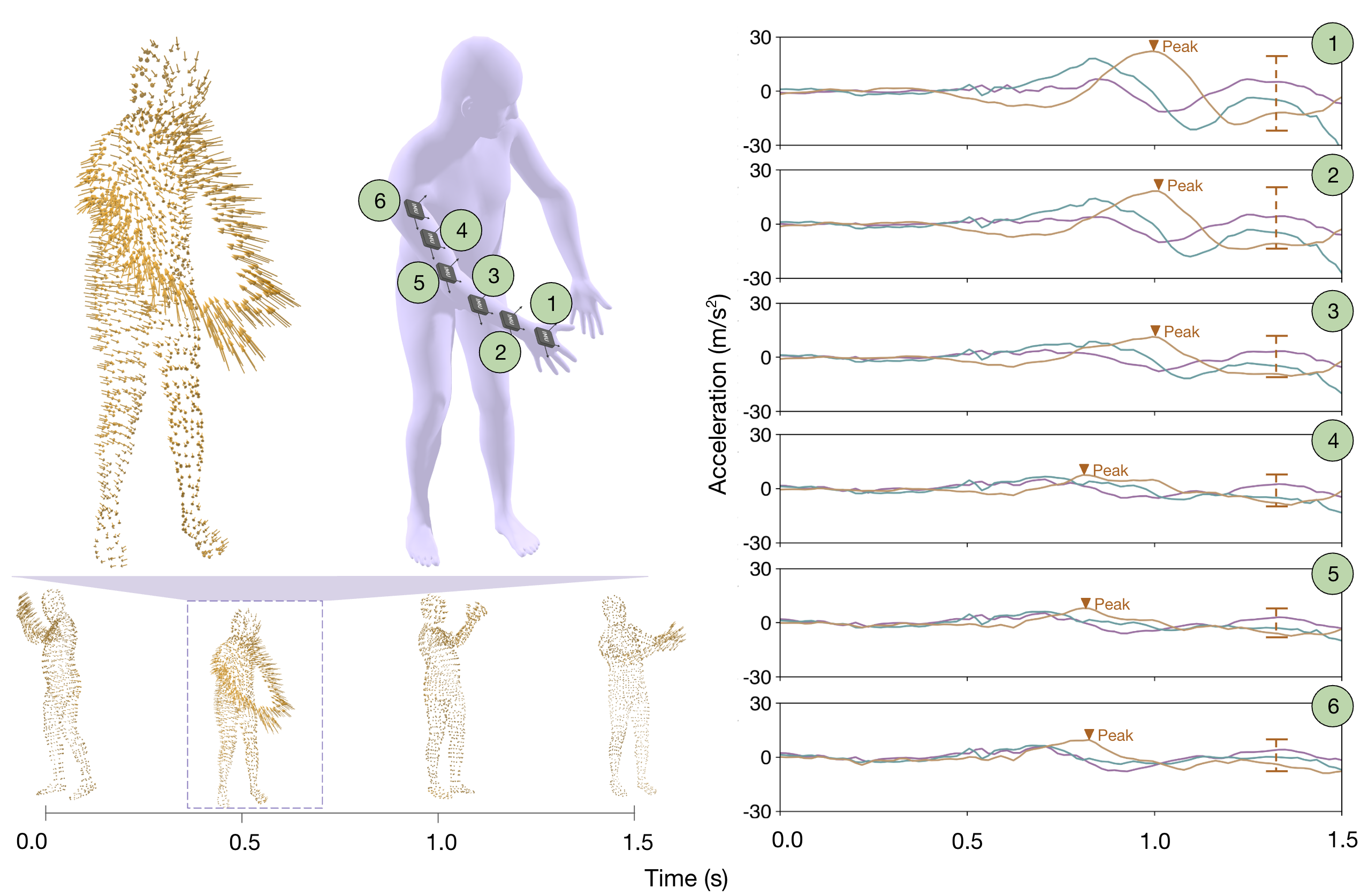}
    \caption{Synthesized acceleration signals across the body surface from a person swinging a golf club from T=0.0s to T=1.5s. The direction and magnitude of acceleration are visualized as arrows (Left Figure). The acceleration selected from 6 ``devices'' on the right arm from hand to shoulder is plotted (Right Figure), corresponding to the six devices on the avatar's right arm. The magnitude and timing of the peak in the acceleration signal at each location are different. \proposed{} models such differences and supports placement and movement of the IMU to different points on the body to give the user flexibility.}
    \Description{Visualization of synthesized acceleration signals on the body during a golf swing motion. Left: The full-body figure represents acceleration vectors (direction and magnitude) from T=0.0s to T=1.5s using gold arrows over a body mesh. Center: An avatar model highlights six IMU device positions labeled 1–6 along the right arm from hand to shoulder. Bottom: Snapshots show the avatar's motion at different time points during the swing. Right: Acceleration graphs for each of the six sensors show time-series data with three-axis acceleration (x, y, z). Key moments, such as peak acceleration, are annotated.}
    \label{fig:system_intuition}
\end{figure*}

However, the inherent assumption in these approaches is that each IMU, while part of a consumer device, is still a specialized device, worn at specific locations. 
Under this assumption, machine learning models have to be trained for specific devices and placements, which limits their practicality.
For example, IMUPoser ~\cite{DBLP:conf/chi/MollynAG0A23} inferred 3D body pose using a few common device placements, such as pants pocket, ear, and wrist. 
The authors built a model that adapted to varying availability of devices and worked on consumer products.
However, a user might want to move the same device to different locations over the course of a day. 
Current models fail to adapt when users choose to carry their phone in a jacket pocket, wear it on an armband while exercising, or mount it on their back as a posture tracker while seated.
For each new location, the ML models need to be retrained, and that severely limits the flexibility of any ML model aimed at ubiquitous human motion sensing.

To bridge this gap, we need to reconsider the potential source of IMU signals for training the models and not remain limited to a set of key locations, but to any point on the human body's surface.
\figref{fig:system_intuition} shows the full-body acceleration of a golf swing captured by a camera-based motion capture system. 
Each arrow in the figure represents the acceleration vector synthesized at that point. 
It also plots the temporal changes in synthesized acceleration for 6 key points on the user's arm.
Signals from devices placed on regions that move together appear to be kinematically related, with some cases observed in previous studies~\cite{hicheur2006intersegmental, wada2014correlation}.
Although sometimes certain parts of the body can move entirely independently, in many other situations, these correlations suggest the existence of an underlying mechanism that connects such signals.
Inspired by this observation, the question arises: \textbf{Can we create a unified model that can take the IMU data captured from any point on the human body surface for human motion sensing?} 

To achieve this vision, there are several technical challenges that need to be addressed.
First, no existing machine learning architecture for IMU data is designed to process input from all possible points on the human body surface.
Existing approaches rely on creating separate feature encoders for each location (\eg \cite{DBLP:conf/sensys/JeyakumarLSS19}) or training together with multiple inputs (\eg~\cite{DBLP:conf/chi/MollynAG0A23}). 
Thus, learning a model for a large number of inputs remains unscalable, especially when the number of possible input locations is infinite. 
Second, there is no existing dataset that includes data from nearly infinite possible IMU locations. 
To our knowledge, the most comprehensive IMU-based dataset yet uses 17 real IMUs~\cite{DBLP:journals/tog/HuangKABHP18}, which is still far from our goal. 
Third, a training framework to learn the relationship between the large number of IMU signals on the body surface and the underlying human motion has not yet been built.

In this paper, we present \textit{\proposed{}}, a novel approach enabling IMU placement over ``Continuous Coordinates'' that maps IMU signals obtained from different locations on the body into a unified feature space defined by the spatial coordinates of the sensor.
\proposed{} models the kinematic relationship between body locations using a self-supervised learning strategy trained on extensive synthetic IMU signals generated from existing motion capture datasets~\cite{DBLP:conf/iccv/MahmoodGTPB19, DBLP:journals/tog/HuangKABHP18, emokine2024, geissinger2020motion, guidolin2022unipd, maurice2019human, palermo2022raw}.
To address the scalability challenge,   
\proposed{} learns to align IMU signals placed flexibly on the body's surface with a constant number of human joint movements.
The unified mechanism for modeling body motion is driven by the insight that signals across the body surface are influenced by the shared kinematic structure connected by the joints of the human body, thus leading to a model of the body motion regardless of the number of input signals and does not need to be retrained when placements change. 

To demonstrate the performance of our system, we conducted a series of evaluations.
In Evaluation~\#1, we collected a custom dataset for IMU data at atypical locations on a user's body and ground truth recorded using camera-based motion capture. 
We demonstrated that our approach generalizes well to all atypical locations that are not covered by prior approaches. 
In Evaluation~\#2, we showed that for typical locations, our approach achieves competitive performance compared to the existing work~\cite{DBLP:conf/cvpr/WouweLFDL24, DBLP:journals/tog/YiZ021, DBLP:conf/cvpr/YiZHSGT022, DBLP:conf/siggraph/Yi0X24}. 
Finally, we demonstrate \proposed{}'s utility in supporting different usage scenarios ranging from allowing users to use atypical device placements, the ability to change device locations depending on the context, and suggesting placement based on the use case.
Thus, \proposed{} provides a human motion sensing model that allows users to wear their devices as they need or prefer.
It also supports users in tracking their bodies for specific applications by allowing them to move devices as needed.

\section{Related Work}
\label{sec:rw}

Our research builds on existing studies that utilize on-body IMUs for human motion sensing.
In this section, we provide a brief overview of these efforts. 
We also surveyed recent advances in other on-body sensors that might augment IMUs for motion sensing in the future. 

\subsection{Body Pose Tracking from IMUs}

Inside-out full-body pose tracking from sparse sensor sets has been intensively studied using different sensing modalities, such as vision~\cite{DBLP:journals/tog/ShiratoriPSSH11,DBLP:conf/eccv/JiangSQFLSH22,DBLP:conf/chi/AhujaMG021}, RFID~\cite{DBLP:conf/huc/JinWYKH18}, pressure~\cite{DBLP:journals/imwut/GaoZRZJWSXJ24}, acoustic~\cite{DBLP:journals/imwut/MahmudLHCJZGZ23}, and electromagnetic-field~\cite{DBLP:conf/mobisys/WhitmirePP19,DBLP:journals/imwut/ArakawaZKGN23}.
Among these, IMU-based solutions are often more deployable, as they are present on many consumer devices.
These solutions learn the mapping between the IMU signals from designated locations on the body to the body pose parameters, which is often represented as the SMPL model~\cite{DBLP:journals/tog/LoperM0PB15}.
For instance, Sparse Inertial Poser~\cite{DBLP:journals/cgf/MarcardRBP17} and Deep Inertial Poser~\cite{DBLP:journals/tog/HuangKABHP18} are pose tracking systems with 6-17 on-body IMUs.
Several other works have been proposed to improve the accuracy of these pose tracking systems~\cite{DBLP:journals/tog/YiZ021,DBLP:conf/cvpr/YiZHSGT022,DBLP:conf/siggraph/Yi0X24,DBLP:conf/cvpr/ZhangXCY0P24}.
Researchers have also explored ways for more practical device placements; for example, IMUPoser~\cite{DBLP:conf/chi/MollynAG0A23} and MobilePoser~\cite{DBLP:conf/uist/XuGHA24} support different combinations of commercial devices, including earbuds, phones, and watches.

Although current systems achieve high accuracy in pose estimation, they rely on placing IMUs on predefined body parts.
This rigidity restricts adaptability, particularly when encountering placements not represented in the training data.
For example, what if a user prefers to wear an anklet instead of a watch?
While recent approaches such as DiffusionPoser~\cite{DBLP:conf/cvpr/WouweLFDL24} attempt to generalize by using diffusion models to infer missing sensor data, they are still confined to 13 predetermined sensor locations.
Consequently, sensor placements outside these predefined points remain unsupported. 
Additionally, the computationally intensive multi-step diffusion process significantly hampers real-time deployment on resource-limited devices.
Crucially, existing research overlooks opportunities to explore the vast and continuous space of the human body for sensor placement, potentially missing configurations that could yield superior accuracy and usability.
In this paper, we explore this opportunity and demonstrate its benefits.

\subsection{Physical Activity Recognition from IMUs}

\begin{figure*}
    \centering
    \includegraphics[width=\linewidth]{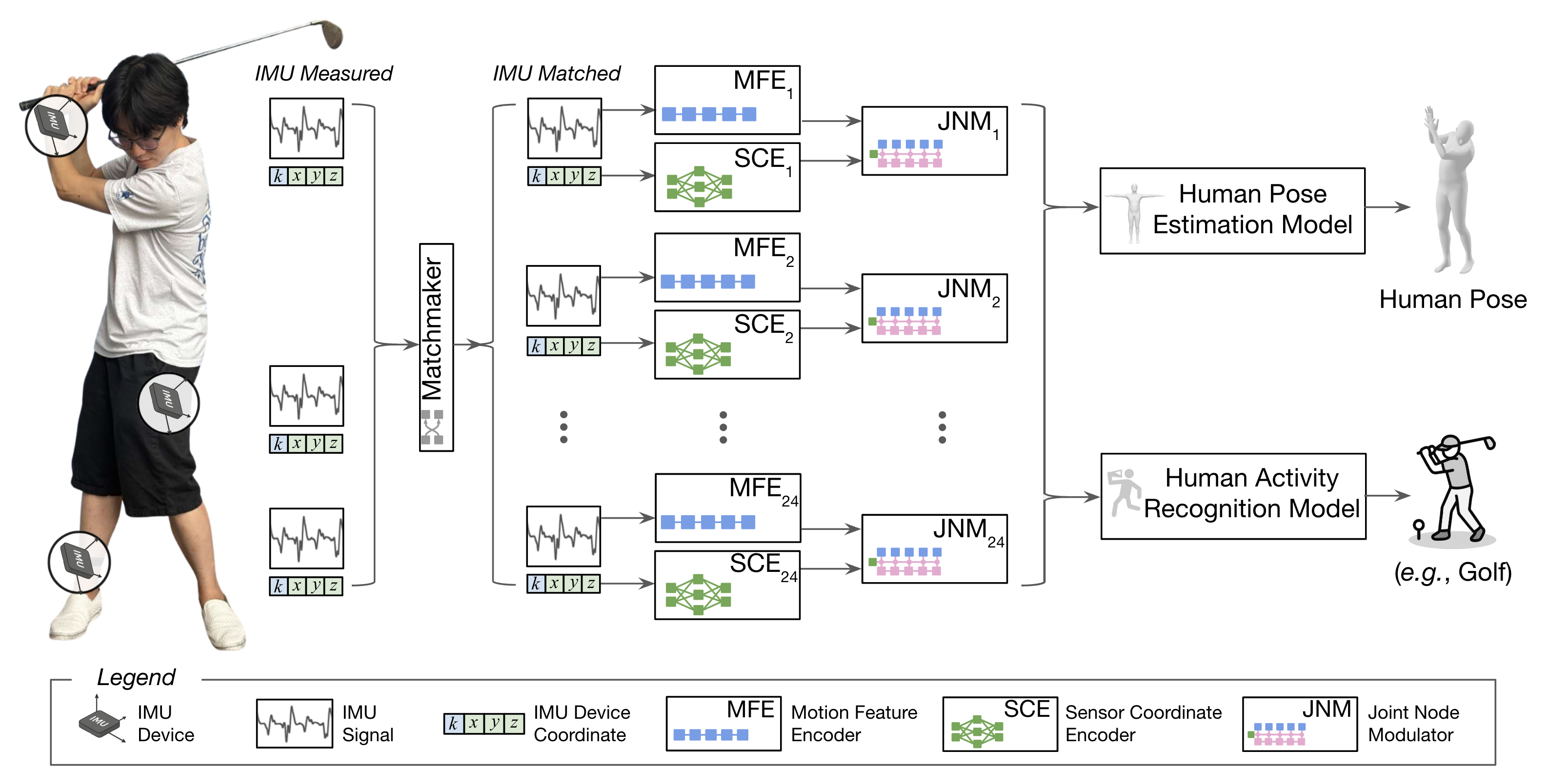}
    \caption{Overview of the \proposed{} system during test time. IMU data is mapped into a placement-adaptive representation, which can be easily tuned for downstream tasks, such as pose estimation and activity recognition. Modules and synthetic IMUs that are only used during training time are omitted from the figure for brevity of presentation.}
    \Description{Overview of the IMUCoCo system at test time. The figure illustrates how IMUCoCo processes IMU sensor data for downstream tasks like human pose estimation and activity recognition. Left: A person with IMU devices placed on various body parts. Each IMU sends raw sensor signals and provides its sensor coordinate (k, x, y, z), which are passed through the system. Center: Matchmaker matches IMUs to the modules based on their placements. Each matched signal is processed through two parallel modules: Motion Feature Encoder (MFE) extracts temporal features. Sensor Coordinate Encoder (SCE) encodes spatial information based on IMU placement. Both outputs are fused via the Joint Node Modulator (JNM). This process is repeated for up to 24 sensors.  Right: Outputs are sent to two downstream models: Human Pose Estimation Model, producing a 3D pose. Human Activity Recognition Model, classifying actions.}
    \label{fig:system_overview}
\end{figure*}

Body motion data can also enable human activity recognition~\cite{DBLP:journals/imwut/ArakawaYMNRDRMC22,Ashry2020} and exercise tracking~\cite{DBLP:conf/bibe2/TianZSWZ21,DBLP:journals/imwut/XiaFAS22}.
Several datasets have been developed that collect sensor data from diverse devices such as smartphones and smartwatches during various daily physical activities, including walking and climbing stairs~\cite{DBLP:journals/sigkdd/KwapiszWM10,DBLP:conf/petra/ReissS12,DBLP:journals/prl/ChavarriagaSCDTMR13}.
Using these datasets, numerous machine learning-based approaches have emerged~\cite{DBLP:journals/csur/BullingBS14,Ramanujam2021,DBLP:journals/imwut/ZhangWCTZG22}.
For example, Müller~\etal~\cite{DBLP:journals/sensors/MullerMAG24} developed a method specifically for exercise tracking using IMUs placed on the wrists and ankles. 
Additionally, Kwon~\etal~\cite{DBLP:journals/imwut/KwonTHGALP20} introduced a technique for synthesizing IMU data from videos, allowing more flexible activity recognition.

However, these methods continue to follow the fixed-placement approach previously discussed in body pose tracking research, relying on a predetermined set of IMU locations.
To overcome this limitation, some studies have investigated location-invariant methods.
Rey~\etal~\cite{DBLP:conf/iswc/ReySL22} proposed a method to transfer sensor data across typical placements (\eg from wrist to ankle). 
SenseHAR~\cite{DBLP:conf/sensys/JeyakumarLSS19} combined multiple sensor types from wearables and phones into a shared latent feature space, enabling downstream activity recognition models to train on unified features. 
Adaimi~\etal~\cite{DBLP:journals/corr/abs-2402-03714} developed a location-invariant model using a large-scale dataset suitable for diverse activities such as driving.
Still, these location-invariant approaches are limited to the sensor placements explicitly captured in their datasets.
In contrast, we demonstrate that \proposed{} can seamlessly integrate into activity recognition pipelines, accommodating a continuous range of sensor placements beyond those predefined locations.

\subsection{Advancements in On-Body Devices}
\label{sec:rw-devices}

Existing on-body sensing research predominantly utilizes common wearable locations, such as glasses, wrists, and thigh pockets.
Beyond these typical placements, researchers have also explored other locations and form factors, such as embedding sensors within everyday accessories~\cite{DBLP:conf/iswc/GemperleKSBM98}, magnetic sensing through rings~\cite{DBLP:journals/cacm/PariziWP22}, multimodal eating detection via necklaces~\cite{DBLP:journals/imwut/ZhangZNXSHA20}, and touch input through belts~\cite{DBLP:conf/chi/DobbelsteinHR15}.
Recent advances in fabrication technologies and robotics have further expanded feasible sensor placements across continuous body surfaces.
For instance, Yu~\etal~\cite{DBLP:conf/chi/YuAMG23} introduced a scarf-shaped device integrated with electrical impedance tomography for activity recognition.
Similarly, SkinMarks~\cite{DBLP:conf/chi/WeigelNOS17} used stretchable electronic tattoos strategically placed on epidermal landmarks, transforming natural skin features and limb movements into interactive touch inputs.
We anticipate that these innovations will facilitate the integration of IMU sensors to capture complete body motion data in the future.
Our proposed system, \proposed{}, can potentially benefit from these advances to enable rapid prototyping of motion sensing systems without requiring additional data collection or extensive model training.

\section{Design of \proposed{}}
\label{sec:proposed}

\subsection{System Overview}

\subsubsection{Insights}
\proposed{}'s architecture aims to enable scalable learning of mapping countless potential placements of IMUs into a tractable space. 
Unlike conventional approaches, \proposed{} enumerates a massive number of possible placements at training time.
For this reason, we must restrict the growth of the architecture size with respect to the number of IMU placements. 

We draw insights from several related fields. Articulated human pose models ~\cite{DBLP:journals/tog/LoperM0PB15} render realistic human body meshes from compact parameters, such as joint rotations and body shapes. 
Techniques such as blend skinning formulate mesh surfaces as weighted bone transformations. 
Inspired by this, \proposed{} is designed to align IMU signals to joint movement representations, limiting the target space to a constant size. 
In addition, coordinate-based or implicit neural representation approaches model signals as a neurally parameterized function of coordinates, demonstrating a powerful ability to perform tasks such as view synthesis ~\cite{mildenhall2021nerf} and compression ~\cite{strumpler2022implicit}.
Leveraging this idea, \proposed{} modulates the IMU signals by their coordinates, effectively encompassing the spatial relationship among various placements without repeatedly increasing the size of the input layers. 

\subsubsection{Architecture}

The \proposed{} maps a variable number of IMU devices placed on the body to a unified space, so that downstream models, such as a pose estimation model, can function without retraining for different IMU placements.
The placements of the devices are represented as spatial coordinates in the 3-dimensional space, formally as $\mathbf{r} = (x, y, z)$.
This spatial coordinate is measured relative to the root of the human body during a standard T-pose and remains unchanged until the IMU device is repositioned or removed from the human body.
Such location information can be easily obtained for users and does not need to be error-free. 
For example, one approach is simply to tap on the corresponding location on a rendered avatar or simply state one of the predefined locations, such as the neck, ankle, or elbow.
We discuss other potential approaches to obtain coordinates in real life in \secref{sec: limitation}.

Inspired by the kinematic trees used for articulated human body models~\cite{DBLP:journals/tog/LoperM0PB15}, \proposed{} breaks the whole body feature space into 24 joint nodes, represented as $\mathbf{z} = \{\mathbf{z}_1,.., \mathbf{z}_{24}\}$, which corresponds to the motion of each joint.
Each of these joint nodes is then processed through a separate pathway that maps the IMU device's signal into the corresponding joint node features.

As shown in \figref{fig:system_overview}, \proposed{}  comprises several modules.
First, each IMU's signal is passed through a Motion Feature Encoder (MFE) to encode the raw IMU data.
At the same time, device placement is encoded using a Sensor Coordinate Encoder (SCE) to derive placement codes that inform the transfer function for the corresponding features into the target joint node feature.
After this step, the Joint Node Modulator (JNM) takes both the extracted IMU features and the placement codes.
It modulates the features to produce a representation that describes the target joint's movement.
To train these modules, the features transferred to the joint node are passed to auxiliary regression tasks using Kinematics Regressors (KRs) to regress to kinematics attributes, including velocity, position, and orientation, as well as a full-body Pose Regressor (PR) to infer the full-body pose. 
The KRs and PR are dropped once the training is completed.
A detailed training architecture and procedure are illustrated in Appendix \ref{appendix:training}.
In the next subsections, we describe each module of \proposed{}, placement adaptation, and applying \proposed{} for downstream tasks in further detail.

\subsection{\proposed{} Modules}

\subsubsection{Motion Feature Encoder (MFE)}

The Motion Feature Encoder (MFE) module encodes the raw IMU input into the IMU feature representations.
This module is analogous to the conventional method of feature extraction for IMU signals.
The MFE module first projects the data from a single IMU's 9 channels into higher dimensions using a linear layer with ReLU activations. 
We implement MFE using an LSTM model following previous studies ~\cite{DBLP:conf/siggraph/Yi0X24,DBLP:journals/tog/HuangKABHP18,DBLP:conf/chi/MollynAG0A23}.
We chose a single-directional LSTM over a bi-directional one to preserve historical information without the limitation of fixed window sizes~\cite{DBLP:conf/cvpr/YiZHSGT022}.
Formally, the MFE module for a joint node is represented by $\mathbf{h} = \text{MFE}(\mathbf{s})$, where $\mathbf{s}$ is one input IMU signal from one point on the body, and $\mathbf{h}$ is the extracted feature.

\subsubsection{Sensor Coordinate Encoder (SCE)}

\begin{figure}
    \centering
    \includegraphics[width=\linewidth]{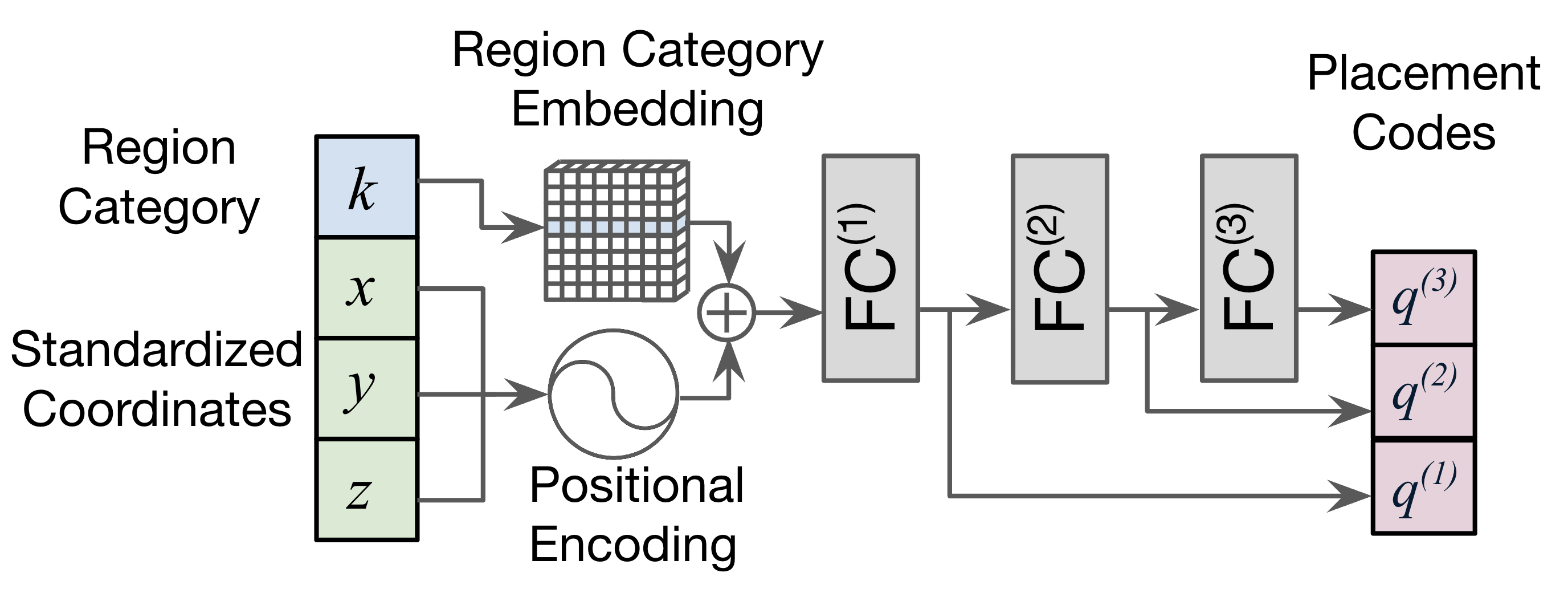}
    \caption{The detailed architecture of the Sensor Coordinate Encoder (SCE) module. The standardized sensor coordinates $(x, y, z)$ are encoded using periodic functions. The sensor region category $(k)$ is encoded using a learnable embedding layer. The concatenated features are passed through fully connected (FC) layers, producing multi-layers of placement codes $\mathbf{q}^{(l)}$ for each layer $l$.}
    \Description{Architecture of the Sensor Coordinate Encoder (SCE) module. This figure shows how spatial information from an IMU is encoded to produce placement codes used for motion understanding. Inputs: The region category k is converted into a learnable region embedding. The standardized 3D coordinates (x, y, z) of the IMU are passed through a positional encoding using periodic functions. Processing: The region embedding and positional encoding are concatenated and passed through a series of three fully connected (FC) layers.}
    \label{fig:sce_module}
\end{figure}

The Sensor Coordinate Encoder (SCE) module encodes the sensor coordinate $r=(x, y, z)$ into placement codes $q$ that instruct the subsequent modules of \proposed{} to adapt to the sensor's placement. 
The detailed structure of SCE is shown in~\figref{fig:sce_module}.
First, we standardize the raw spatial coordinates to have the origin at the target joint location $r_j$, divided by the range of all vertex spatial coordinates.
Formally, this is represented by $r_{st} = (r - r_j)/(r_{max} - r_{min})$
, where $r_{max}$ and $r_{min}$ are obtained by taking the maximum and minimum values of each of the three axes on the coordinates of all vertices. 
Previous research has shown that applying positional encoding helps extract high-frequency changes along coordinates~\cite{mildenhall2021nerf}.
Following this, the standardized coordinate $\mathbf{r}_{st}$ is passed through a positional encoding using periodic functions.
This mapping is defined as:
\begin{equation}
\phi_f(\mathbf{r}_{st}) = \left[ \sin(2 \pi f \cdot \mathbf{r}_{st}), \cos(2 \pi f \cdot \mathbf{r}_{st}) \right]
\end{equation}
, where $f$ represents the frequency bands in increasing powers of $2$, \ie $f = 2^p$ for $p = 0, 1, \dots, n_{freq}-1$.
This encoding transforms $r_{st}$ into a high-dimensional feature space that provides multi-scale spatial information.

Additionally, we partition the human body surface into 24 regions based on joint positions. 
Our intuition is that the region where the sensor is placed provides a semantic meaning that is useful to determine the transference as well.
A detailed illustration of the partition is provided in Appendix \ref{appendix:training}.
Thus, for each IMU sensor, we categorize its placement region into one of the 24 regions, denoted as $k$, based on the provided coordinate $r$. 
Subsequently, we encode the category of regions $k$ corresponding to the spatial coordinate using a learnable embedding layer, denoted as $Emb$.
The positional encodings and the region embedding are concatenated to form the input feature vector, denoted as $\mathbf{q}^{(0)} = \left[ \phi(r_{st}), \text{Emb}(k) \right]$, to the subsequent $L$ fully-connected (FC) layers.

The first FC layer takes input from $\mathbf{q}_0$, while subsequent fully connected layers produce placement codes.
Formally, for layer $l$:
\begin{align}
\mathbf{q}^{(l)} &= \text{FC}^{(l)}\left(\mathbf{q}^{(l-1)}\right)
\end{align}
, where $\mathbf{q}^{(l)} = (\mathbf{\gamma}^{(l)}, \mathbf{\beta}^{(l)})$ are the placement codes that subsequently inform the modulation of the motion features based on placement.
Finally, the output of the SCE module is a list of placement codes $\mathbf{q} = \{\mathbf{q}^{(1)}, \dots, \mathbf{q}^{(L)}\}$, one for each MLP layer.

\subsubsection{Joint Node Modulator (JNM)}

The Joint Node Modulator (JNM) modulates the features obtained from the MFE module to the features representing the corresponding joint kinematics on the body based on $q$, the placement codes generated from the SCE module. 
Similar to MFE, we adopt an LSTM-based structure to encode the temporal information of the motion feature corresponding to each joint node.
For each layer, we applied Feature-wise Linear Modulation (FiLM)~\cite{perez2018film} to the LSTM outputs to allow modulating the motion features based on the inferred placement codes.
The input to the first layer is just the output of the corresponding MFE module $\mathbf{z}^{(0)} = \mathbf{h}$.
Formally, for layer $l$ in the JNM module,
\begin{equation}
\mathbf{z}^{(l)} = \text{LSTM}^{(l)} \left( \mathbf{\gamma}^{(l)} \odot \mathbf{z}^{(l-1)} + \mathbf{\beta}^{(l)} \right)
\end{equation}
Finally, the output of the JNM module is the last layer's output $z^{(L)}$.

\subsubsection{Kinematics Regressor (KR) and Pose Regressor (PR)}
We used regressions to kinematic attributes as an auxiliary task to learn useful representations of cross-placement IMU signals.
To achieve this, we used two linear layers with ReLU activation as a Kinematics Regressor (KR). 
We used this straightforward architecture for KR primarily to enforce the quality of representations learned only from the previous modules instead of letting a complex model excessively compensate for the performance of the regression task.
For each joint node, we used five joint-level KR modules to predict the joint's velocity, position, local orientation, global orientation, and the body root's velocity.
In addition, we used one body-level Pose Regressor(PR) module, with the same architecture as KR, that takes the features from all 24 joints and together regresses to the full-body pose.
Note that the KRs and PR are used as auxiliary regressors only and will be removed from the model once it is trained. 

\subsubsection{Matchmaker}
When \proposed{} is supplied with a variable number of IMUs, the Matchmaker module dynamically allocates joint nodes to the optimal IMU device based on a loss map. Specifically, after training, for each joint node, we iterate through all vertices on the body, assuming that an IMU is placed at each vertex, and compute the loss values for transferring information to the target joint node. This process is repeated for all 24 joints, producing a loss table $\mathbf{M} \in \mathbb{R}^{24 \times V}$, where $V$ is the total number of vertices in the articulated pose model, and $\mathbf{M}(j,v)$ represents the loss value of a virtual IMU placed at the vertex $v$ when transferred to the target joint node $j$. 
With the constructed loss table, each of the 24 joint nodes will be assigned to one IMU device that gives the lowest loss.
Formally, given a set $\mathcal{D}$ of IMU devices attached to the body, where the IMU $d$ is located at the coordinate $\mathbf{r}_d$, the optimally assigned IMU for the joint node $j$ is computed as $d_j^* = \argmin_{d \in \mathcal{D}} \mathbf{M}(j, v_{\mathbf{r}_d})$, where $v_{\mathbf{r}_d}$ denotes the nearest vertex in the loss table to the provided coordinate $\mathbf{r}_d$. For a newly attached or moved device, we use its coordinates to query the table and retrieve 24 loss values, and update the assignment accordingly.
Note that this module is only used during test time.

\subsection{Training Process}

As mentioned in the introduction, the absence of a dedicated dataset for IMU data on continuous body coordinates presents a technical challenge. We introduce an approach to synthesizing virtual IMU data across the entire body mesh, expanding beyond the joint-only focus of previous research.
This section explains the approach, followed by details of our training approach.

\subsubsection{Virtual IMU Synthesis}

Training \proposed{} requires IMU data sampled from all over the body surface, for which relying on real IMU data is not feasible.
We synthesized our IMU data based on human pose datasets, including AMASS~\cite{DBLP:conf/iccv/MahmoodGTPB19}, DIP-IMU~\cite{DBLP:journals/tog/HuangKABHP18}, and other XSens-based datasets~\cite{emokine2024, geissinger2020motion, guidolin2022unipd, maurice2019human, palermo2022raw}.
From these datasets, we obtained full body pose and used the SMPL model~\cite{DBLP:journals/tog/LoperM0PB15} for forward kinematics to determine the positions of joints and mesh. 

To calculate the acceleration, we applied the second derivative of the positions.
Existing work typically simplifies bone orientation to synthesize IMU orientation (\eg ~\cite{DBLP:conf/chi/MollynAG0A23}).
While this approach may function well when the IMU is attached to areas that do not deform, such as the middle of the upper arm, it is inaccurate for areas that deform, such as abdominal regions and areas close to each joint (\eg the elbow).
Thus, we calculate the orientation based on the faces of the mesh. 
These virtual IMUs are also virtually calibrated using a T-pose~\cite{DBLP:journals/tog/YiZ021}
We include a more detailed illustration in \appref{app:system-virtual}.

\subsubsection{Loss Functions}

We used a combination of loss functions to establish learning feature representations suitable for the downstream tasks. 
First, we used kinematic loss $\mathcal{L}_{\text{kinematic}}$ for multiple kinematic attributes, including velocity, root velocity, position, global orientation, and local orientation, to facilitate the model in extracting kinematic quantities.
For velocity, position, global orientation, and local orientation, we used Mean Squared Error (MSE) loss. 
For root velocity, we used multi-frame losses at consecutive 1, 3, 9, and 27 frames ~\cite{DBLP:journals/tog/YiZ021}.
Second, we used the full-body pose loss $\mathcal{L}_{\text{pose}}$ in the global frame, also implemented as an MSE loss, to encourage the model to coordinate organically with the representation from different joint nodes.
Note that this fully-body pose loss can only be calculated when all the joint nodes have completed their forward passes, which, if done at the same time, can add an excessive memory burden during training.
We used a buffered approach to resolve this, as described in~\appref{appendix:training}.
Third, we used an alignment loss $\mathcal{L}_{\text{align}}$ based on cosine similarity to encourage \proposed{} to produce a feature representation for the sampled mesh synthetic IMUs similar to the representation for the joint synthetic IMU.
Formally, the training loss can be represented as:
\begin{equation}
\begin{aligned}
\mathcal{L}_j (\mathbf{z}_{j}) &= \lambda_{\text{kinematic}} \mathcal{L}_{\text{kinematic}}(KR(\mathbf{z}_{j}), \mathbf{K}_{j\text{GT}}) \\
                      &\quad + \lambda_{\text{pose}} \mathcal{L}_{\text{pose}}(PR(\mathbf{z}_{j}, \mathbf{z}_{\text{buffer}}), \mathbf{P}_\text{GT}) \\
                      &\quad + \lambda_{\text{align}} \mathcal{L}_{\text{align}}(\mathbf{z}_{j}, \mathbf{z}_{\text{ref}})
\end{aligned}
\end{equation}
, where $\mathbf{K}_{j\text{GT}}$ are the true kinematics of the joint, $\mathbf{P}_\text{GT}$ is the ground truth pose, $\mathbf{z}_{j}$ is the representation obtained from the mesh IMU, $\mathbf{z}_\text{ref}$ is the reference representation obtained from joint virtual IMU, and $\mathbf{z}_\text{buffer}$ is the buffered representations initially filled with the joint virtual IMU representations while gradually replaced by the mesh IMU representations, and $\lambda_{\text{kinematic}}, \lambda_{\text{pose}}, \lambda_{\text{align}}$ are hyperparameters for weighing the losses. 

\subsubsection{Training Setup}
\label{section:training}

Training such a model consumes a large dataset, and we designed a training scheme to minimize the computing resources needed. 
Our training process is split into two phases. 
In phase one, we train the \proposed{} model only with the 24 joint virtual IMUs using kinematic loss and pose loss until convergence. In phase two, with the warmed-up model, we sampled mesh IMUs and trained them with kinematic and pose loss while aligning them to joint IMUs.
A more detailed procedure is illustrated in Appendix \ref{appendix:training}.
Overall, we trained our model on one L40S GPU, which has 48GB CUDA VRAM, for 200 hours, including 90 hours of training in the joint-only phase and 110 hours of training in the end-to-end phase. 
While training \proposed{}  is relatively prolonged, mainly due to the need to consume a large number of samples on the human body surface, this model is still very lightweight with only 23M parameters and feasible for inference without excessive computing.

\subsection{Applying \proposed{} to Downstream Tasks}
\label{subsection:downstream_task}

The features extracted from \proposed{} enable the downstream model, such as a pose estimation model, to process without worrying about the number or locations of IMU devices.
To train a downstream task, one can freeze the \proposed{} model and feed the IMU data from the devices existing in this dataset for the task. During training, each provided IMU went through the placement adaptation process and was matched to the joint node to produce the extracted representations.
For our paper, we explored pose estimation and activity recognition as our downstream tasks.

For pose estimation, we mainly adopt the DynaIP~\cite{zhang2024dynamic} architecture, as a state-of-the-art pose estimation model using 6-IMU inputs.
As DynaIP itself does not give translation estimation, we adopt the translation estimation module from TransPose~\cite{DBLP:journals/tog/YiZ021}. We use the abbreviation DTP (DynaIP with TransPose) to denote this pose estimation model. For convenience of presentation, in the evaluation section, we use \proposed{} to refer to \proposed{} + DTP in the context of pose estimation. 
We then trained a pose estimator using the conventional 6 IMUs (pelvis, head, left lower arm, right lower arm, left lower leg, right lower leg) from AMASS, DIP-IMU, and XSens dataset ~\cite{zhang2024dynamic}. 

For activity recognition, we used a Spatial-Temporal Graph Convolutional Neural Network (ST-GCN) targeted for skeleton-based activity recognition ~\cite{yan2018spatial}. We then trained our activity recognition model using the 3 IMUs (wrist, pocket, ear) using custom datasets that we collected ourselves. 
For convenience of presentation, in the evaluation section, we use \proposed{} to refer to \proposed{} + ST-GCN in the context of activity recognition. 
Both models are trained when freezing the \proposed{} model and only using the IMU data at the different locations provided in the dataset. 
We conducted separate ablation experiments that verified that the improved performance is attributable to \proposed{} rather than the downstream models.

\section{Evaluation}
\subsection{Evaluation Overview}

We conducted two studies to evaluate the following hypotheses:
\begin{enumerate}
    \item \proposed{} achieves consistent body motion sensing performance when IMUs are placed at atypical locations.
    \item \proposed{} delivers comparable body motion sensing performance to existing systems when IMUs are placed at typical locations.
\end{enumerate}
For Evaluation~\#1, we used our newly collected custom dataset to assess performance in body pose tracking and activity recognition across sensor placements.
For Evaluation~\#2, we utilized multiple existing datasets to evaluate body pose tracking capabilities.

\subsection{Evaluation~\#1: Motion Sensing at Atypical Locations}

\subsubsection{Data Collection}

\begin{figure}
    \centering
    \includegraphics[width=\linewidth]{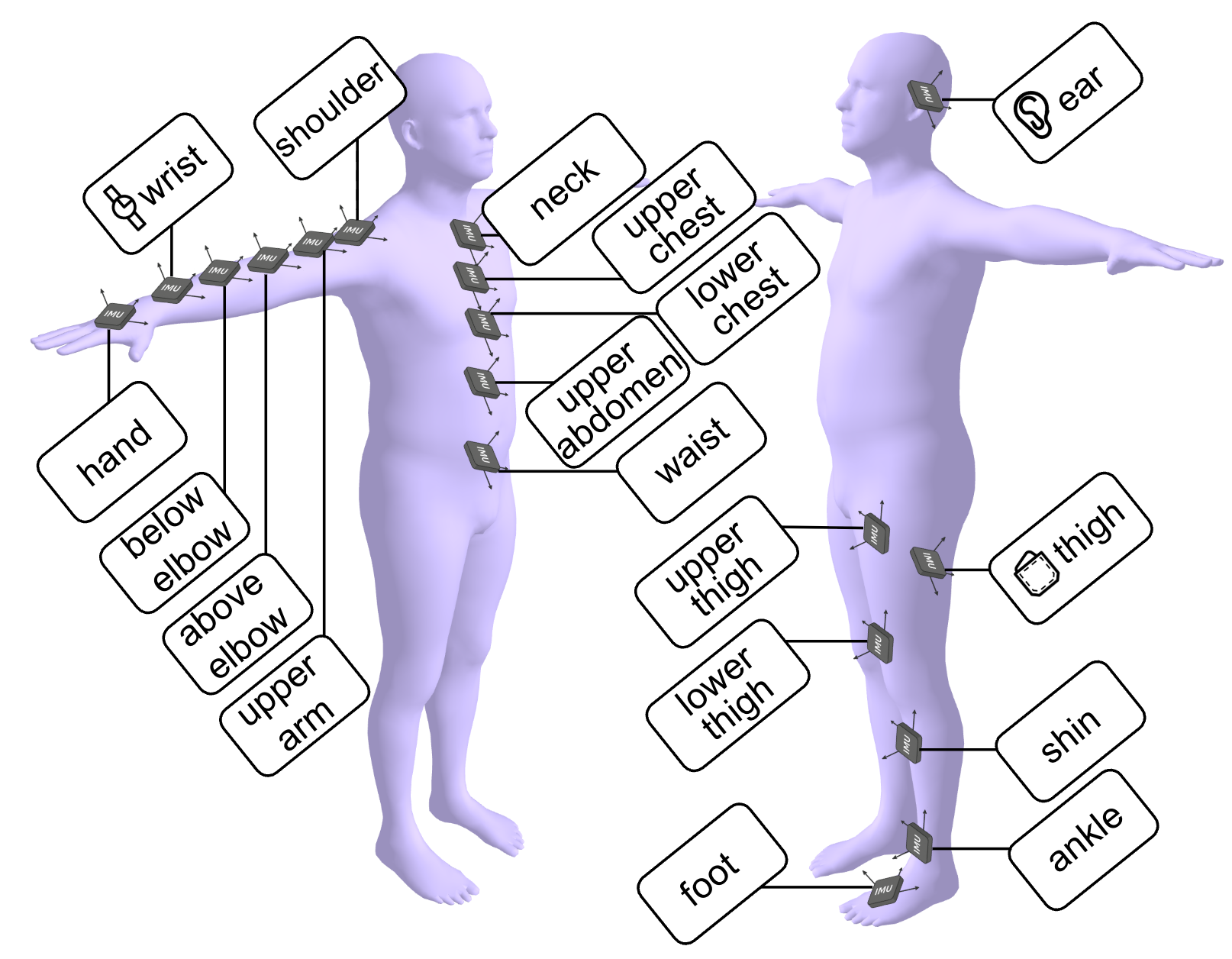}
    \caption{Illustration of the dense IMU placement configurations used in Evaluation~\#1 shown on a right-handed person. We started with the standard 3 IMU placement (wrist, thigh, ear) to mimic the most popular consumer devices~\cite{DBLP:conf/chi/MollynAG0A23}. Additionally, we placed a dense set of IMUs on one of the three body sections (arm, leg, or torso). On the arm, we chose the shoulder, the middle of the upper arm, the upper arm just above the elbow, the forearm just below the elbow, and the hand. On the leg, we chose the upper thigh, lower thigh, shin, ankle, and foot. On the torso, we chose the neck, upper chest, lower chest, upper abdomen, and waist. For a left-handed person, the placements were mirrored accordingly.}
    \Description{Visualization of dense IMU placement configurations used in Evaluation #1 on a right-handed person. The figure shows a full-body avatar with IMUs placed on a comprehensive set of body regions to evaluate pose and motion sensing performance. Standard 3 IMU placements include: Wrist, thigh pocket, and ear, representing common consumer device positions. Dense placements by body section: Arm: hand, below elbow, above elbow, upper arm, and shoulder. Leg: upper thigh, lower thigh, shin, ankle, and foot. Torso: neck, upper chest, lower chest, upper abdomen, and waist.}
    \label{fig:study1-placement}
\end{figure}

Due to the absence of datasets with dense IMU placement, we collected our own custom dataset to understand \proposed{}'s performance at fine-grained placement variations.
We utilized 8 Apple Watches (Series 7 or newer) with a custom data collection application that we implemented to record IMU data at 50~Hz. 
We collected motion capture data using the OptiTrack system.

\begin{figure*}[htbp]
    \centering
    \begin{subfigure}[b]{0.675\textwidth}
        \centering
        \includegraphics[width=\linewidth]{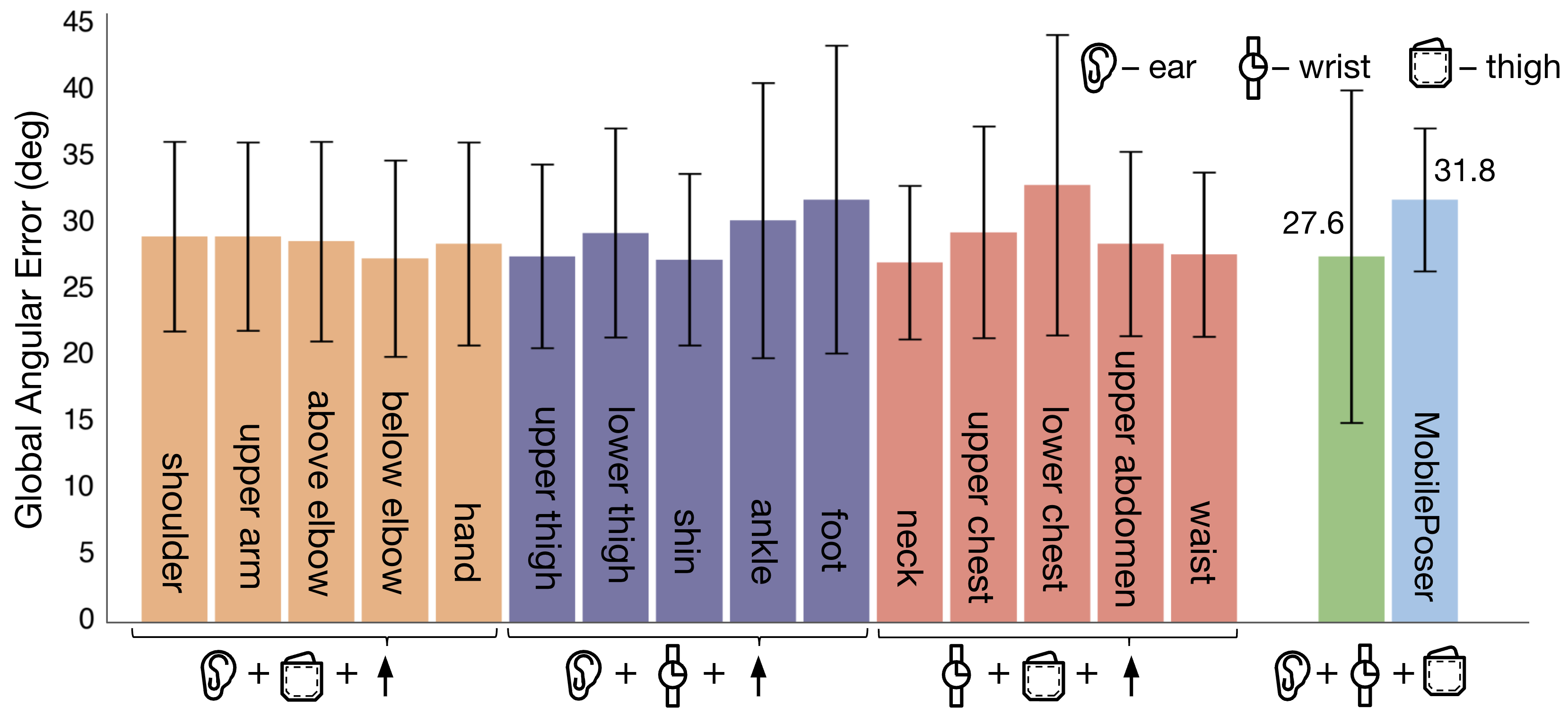}
        \caption{}
         \label{fig:study1-combined-pose}
    \end{subfigure}%
    \hfill
    \begin{subfigure}[b]{0.315\textwidth}
        \centering
        \includegraphics[width=\linewidth]{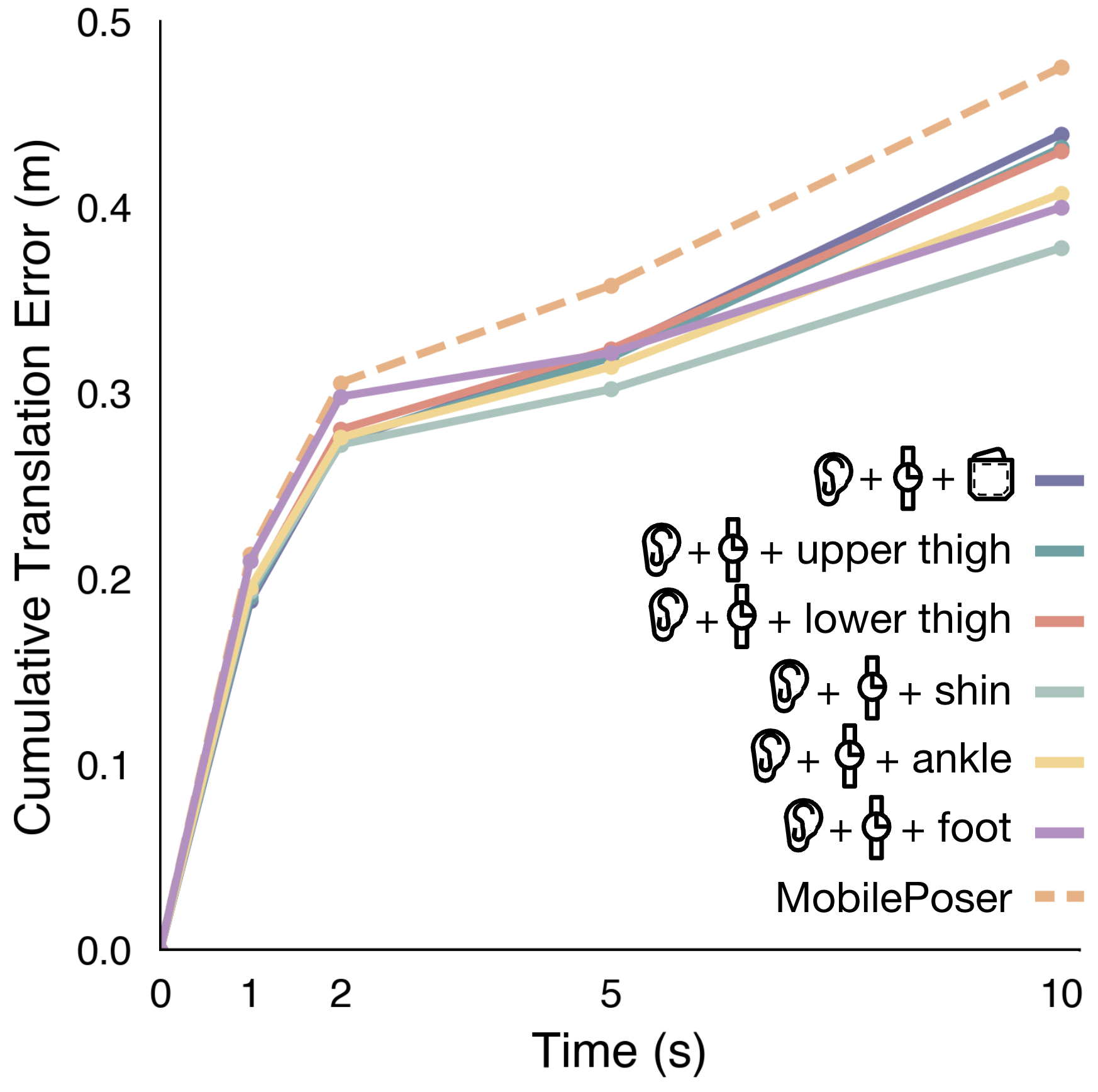}
        \caption{}
         \label{fig:study1-combined-tran}
    \end{subfigure}
    \caption{(a) Pose tracking Global Angular Error (GAE) is measured at different sensor placements, averaged over all activities. \proposed{} achieves lower error than the state-of-the-art~\cite{DBLP:conf/uist/XuGHA24} when IMUs are placed at typical positions (bar graph on the right in (a)). A similar effect is seen at all atypical positions (\ie bars on the left). Error bars show standard deviation. (b) The same effect was seen for cumulative translation error using different sensor placements for the lower body. \proposed{} achieves consistently lower translation error than MobilePoser over time. The effect remains the same as the participant changes the device location along their leg. }
    \Description{Evaluation of IMUCoCo's performance across different IMU placements. (a) The bar chart shows Global Angular Error (GAE) in degrees for different IMU placements during pose tracking, averaged across all activities. Bars are grouped by body region (arm: orange, leg: blue, torso: red, and green for the standard three-sensor setup). IMUCoCo achieves low angular error across all placements, with performance remaining robust even at non-standard positions. Standard placement (green bar) achieves an average error of 27.6°, lower than most individual placements. (b) The line graph plots Cumulative Translation Error (in meters) over time (0–10s) for different IMU placements along the leg. IMUCoCo consistently outperforms the baseline (MobilePoser, dashed orange line) regardless of exact placement (upper thigh, lower thigh, shin, ankle, foot), indicating stable translation accuracy under various conditions.}
    \label{fig:study1-combined}
\end{figure*}

We recruited 12 participants from our institution (8 males, 4 females; age range 23-33; 1 left-handed, 11 right-handed) and collected their motion capture data.
Participants wore a mocap suit with optical markers, followed by skeleton calibration in the OptiTrack system.
The participants also wore watches at three typical locations (ear, wrist, and pocket).
Additionally, we placed five more devices on one of the three dense placement configurations: arm, leg, and torso, as illustrated in~\figref{fig:study1-placement}.
Participants then performed jumps to align timestamps across all IMU devices.
For each placement configuration, participants performed a set of predefined activities inspired by previous research~\cite{DBLP:conf/sensys/JeyakumarLSS19}, designed to encompass diverse body movements: walking, running, vacuuming, watching television, drinking water, table cleaning, golf swing, shot put, and squats.
Before each activity, the participants performed a T-pose for calibration purposes~\cite{DBLP:journals/tog/YiZ021}, with each activity lasting 30 seconds. 
Participants rested between placement configurations as needed, and the entire data collection process required approximately one hour per participant.
All activities were video-recorded with timestamps for subsequent annotation.
Following data collection, we extracted the OptiTrack body pose data, synchronized it with all IMU sensor readings, and annotated the dataset with activity labels.
Overall, our collected dataset consists of 3 hours of body motion data in total. 
All studies are approved by our institution's IRB.

\subsubsection{Body Pose Tracking}
We first evaluated the body pose tracking performance of \proposed{} by comparing different sensor placements, specifically examining how error metrics vary as the IMU sensor is progressively moved along a body part (\ie arm, leg, or torso).
Consistent with prior research~\cite{DBLP:conf/cvpr/WouweLFDL24}, we employ Global Angular Error (GAE)
\footnote{We followed the procedure with DiffusionPoser~\cite{DBLP:conf/cvpr/WouweLFDL24}, where we ignored root, wrists, fingers, and toes joints and excluded them from averaging the final error calculations. For this, our reported error values may appear greater than their originally reported ones.} that serves as our primary metric for pose estimation quality, which quantifies the rotational discrepancy between the ground truth and the reconstructed global segment orientations. 
It is important to note that state-of-the-art systems, such as MobilePoser~\cite{DBLP:conf/uist/XuGHA24}, do not support IMUs positioned at unconventional locations.
Therefore, to keep the comparison fair, we compare \proposed{} and previous work for sensors placed in standard locations.
For atypical positions, we compare the performance within \proposed{}'s output.



\begin{figure*}
    \centering
    \includegraphics[width=\linewidth]{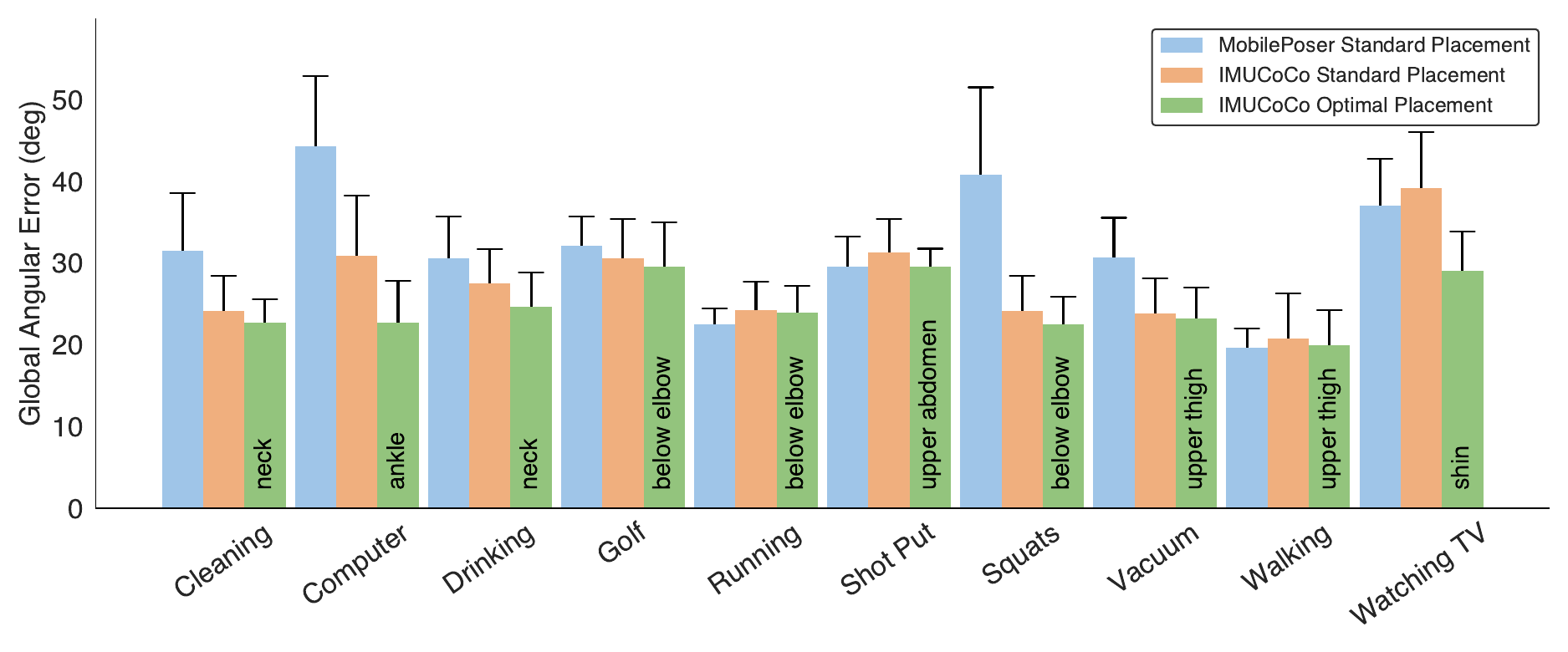}
    \caption{Comparison of Global Angular Error (GAE) when the sensor is placed at Standard Placements (wrist, pocket, ear) and when one of the three sensors is moved to an optimal location. For example, for the Golf activity, the model generates a better pose (\ie lower error) when the IMU on the wrist is moved below the elbow. The Figure also shows MobilePoser's GAE for the standard placement. Error bars show standard deviation.}
    \Description{This bar graph shows Global Angular Error (GAE) for multiple activities under three conditions: MobilePoser Standard Placement (blue): wrist, pocket, ear. IMUCoCo Standard Placement (orange): same sensor positions. IMUCoCo Optimal Placement (green): one of the sensors is repositioned to a better location (\eg below elbow, shin, etc.). Each activity (\ie Cleaning, Computer use, Golf, Squats, Watching TV) shows three bars indicating the relative error of each method. Across nearly all tasks, IMUCoCo (both standard and optimal) performs better than MobilePoser, with the optimal IMU placement yielding the lowest angular error in most cases.}
    \label{fig:study1-pose-location-result}
\end{figure*}

These comparisons are summarized in \figref{fig:study1-combined-pose}. 
Overall, ~\proposed{} showed consistency in performance as the IMU device moves along the arm, leg, and torso regions.
In particular, we observed only slight variation as the IMU is repositioned along the arm.
Slightly higher errors appear when the ear sensor is moved to the lower chest.
We attribute this mainly to the flexibility of clothing, where the physical IMU often folds or shifts together with the fabric, introducing additional motion artifacts that are not representative of true body movement.
In addition, ~\proposed{} demonstrated superior performance (GAE = \ProposedOurDatasetStandardGAE \textdegree) in pose estimation than MobilePoser~\cite{DBLP:conf/uist/XuGHA24} (GAE = \MobilePoserOurDatasetStandardGAE \textdegree) ($p < .001$).
Similarly, ~\proposed{} showed resistance in translation estimation as the user moves the leg IMU from thigh pockets to other areas across the leg, as shown in \figref{fig:study1-combined-tran}.
It is important to note here that, unlike MobilePoser, \proposed{} or DTP are not specifically trained or fine-tuned with the IMU at these locations (wrist, thigh pocket, ear), and thus even the standard placement relies on ~\proposed{}'s capability to transfer input IMU signals to the DTP model that were trained using the 6-IMU configuration.

We then measured the pose estimation performance from ~\proposed{} using standard placement (wrist, thigh pocket, ear) to the optimal placement for each type of activity. \figref{fig:study1-pose-location-result} summarizes the result. 
We see that selecting the best placement for IMUCoCo enables greater tracking performance (GAE = 24.8\textdegree) compared to the standard location (GAE = 27.6\textdegree) ($p < .001$). 
For some specific activities, changing the placement appears to have more impact on the pose estimation accuracy. 
For instance, golf swings are better captured by moving the IMU from the wrist to below the elbow (or the forearm). 
For computer work, moving the IMU from the thigh pocket to the ankle produces less error (GAE=22.8\textdegree) than the standard location of keeping the sensor in the pocket (GAE=30.9\textdegree).
We examined the inferred pose difference and observed that the primary difference is the angle of the hip with respect to the ground when sitting.
We consider that this difference arises mainly due to the more rigid nature of the lower leg clothing and bones, compared to the IMU in the upper thigh pocket, which introduces additional flexibility when sitting. 
Thus, if a user wants to better capture a specific type of posture, \proposed{} can recommend and support optimal sensor placement.

\begin{figure}
    \centering
    \includegraphics[width=\linewidth]{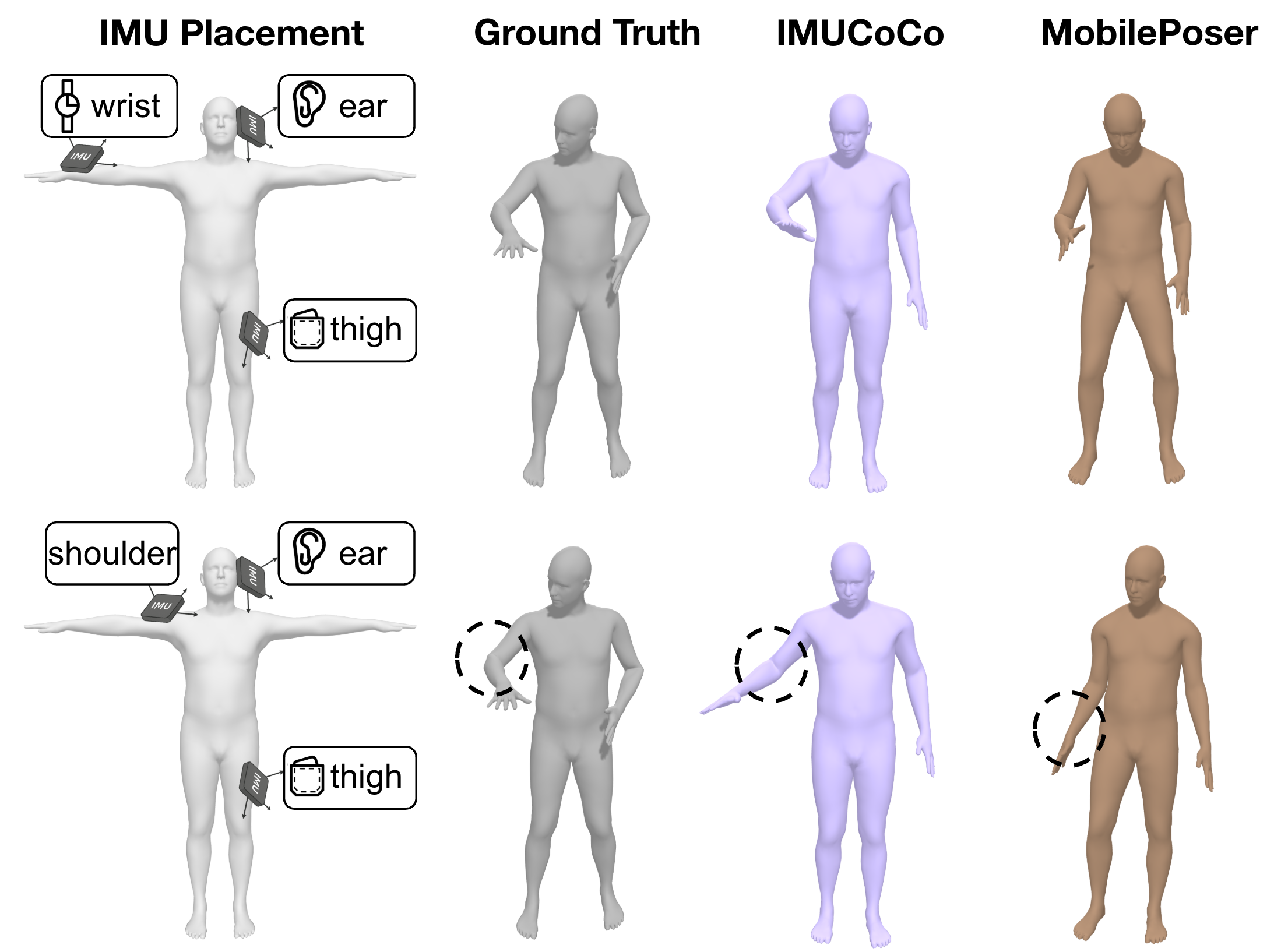}
    \caption{A demonstration of pose estimation when the sensor on the arm is moved, \ie from the wrist (top row) to just below the shoulder (bottom row).  When the user performs a sweeping motion (ground truth), IMUCoCo is able to adapt to the new IMU placement and infer the upper arm motion accurately (bottom row, third column). MobilePoser still assumes the input was at the wrist and incorrectly only slightly raises the forearm (bottom row, fourth column).}
    \Description{Qualitative comparison of pose estimation when the IMU placement on the arm is changed. The figure compares pose estimation results for two IMU placements: Top row: IMU placed on the wrist. Bottom row: IMU moved to the shoulder (just below). Each row shows: IMU Placement illustration (leftmost column). Ground Truth pose (second column). IMUCoCo output (third column). MobilePoser output (rightmost column). When the IMU is at the wrist (top row), both IMUCoCo and MobilePoser estimate the pose reasonably well. However, when the IMU is moved to the shoulder (bottom row), IMUCoCo correctly adapts and infers the lifted upper arm, closely matching the ground truth. In contrast, MobilePoser fails to adjust and assumes wrist-based input, producing an inaccurate pose with only slight forearm movement.}
    \label{fig:study1-example}
\end{figure}

We also evaluated the pose estimation results with different sensor placements in more detail to understand the sources of errors.
Consistent with our hypothesis, ~\proposed{} effectively adapts the sensor based on its placement coordinate, and renders the pose reasonably based on the sensor placement. 
\figref{fig:study1-example} shows an example motion of sweeping to clean a table, which primarily involves arm movements while leaning forward. 
As the IMU is repositioned from the wrist to the upper arm (below the shoulder), ~\proposed{} can adapt the signal and accurately infer the upper arm motion. 
We attempted to retrieve the pose estimation from MobilePoser~\cite{DBLP:conf/uist/XuGHA24}, which is not designed to take IMU input from the upper arm.
As expected, the state-of-the-art that is not designed to adapt to new sensor placements did not render the pose accurately.
It is important to note that ~\proposed{} still lacks sensor input from the lower arm and can only provide its best prediction by naturally extending the lower arm.
In this case, some error in lower arm tracking is anticipated.
Based on these observations, we conclude that understanding tracking confidence for all body parts, given specific sensor placements, and controlling the level of model hallucination is essential for critical applications.

\subsubsection{Body Activity Recognition}
\begin{figure}
    \centering
    \includegraphics[width=\linewidth]{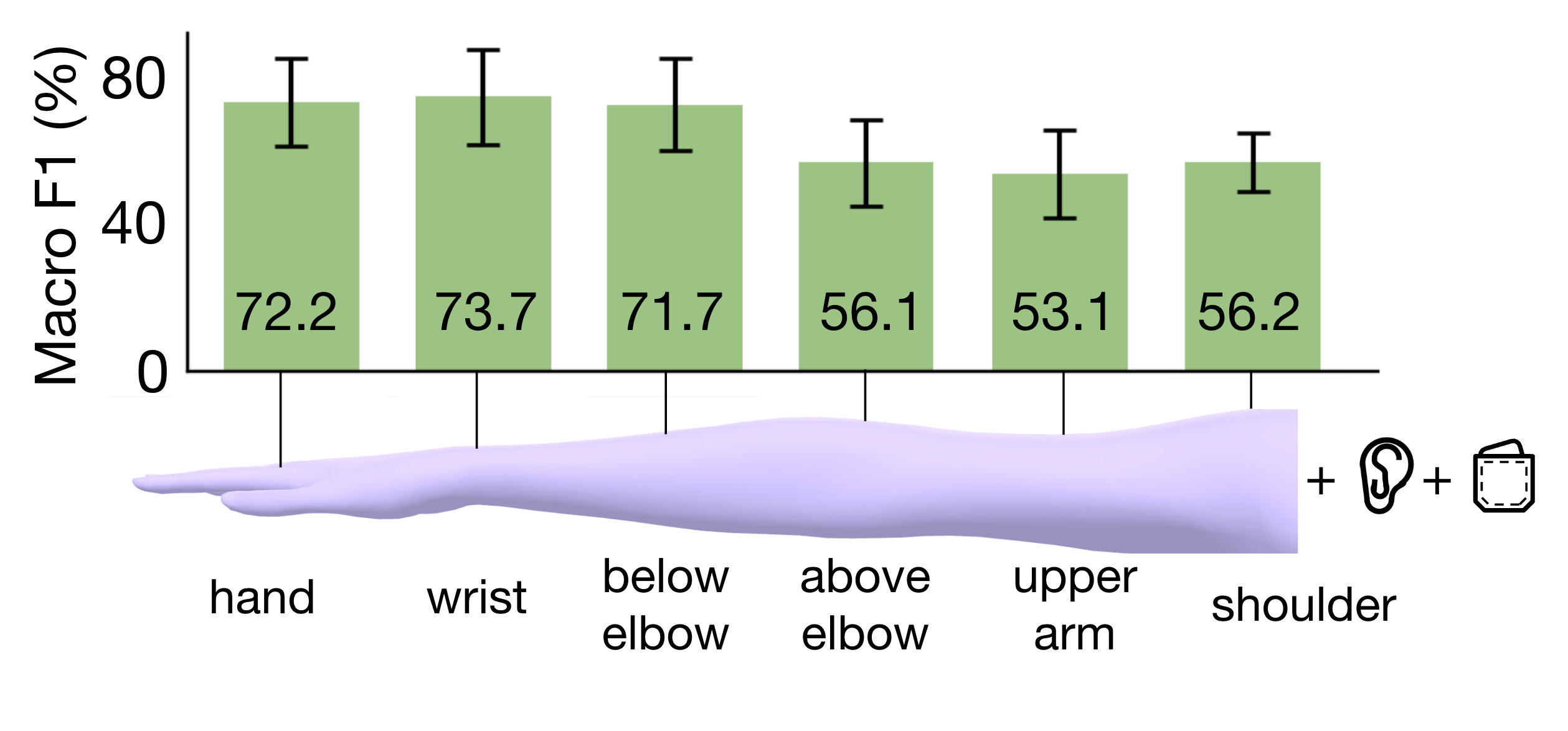}
    \caption{Activity recognition results in macro F1-score across 10 activities with 6-fold cross validation on our custom dataset when placing the arm IMU at different locations. Higher is better. Error bars show standard deviation.}
    \Description{The bar graph shows F1-scores for 10 activities using 6-fold cross-validation, with the arm IMU placed at six different locations (from left to right along the arm in the illustration below the bars). The sensor configuration includes an additional IMU on the ear and one on the pocket (icons shown on the right).}
    \label{fig:study1-activity}
\end{figure}

We divided our 12 participants into six groups of 2 and performed a 6-fold cross-validation to examine the activity recognition performance of our approach. 
\figref{fig:study1-activity} summarizes the results of the averaged macro F1-score from 6-fold cross validation.
Overall, using the three standard placements, \proposed{} achieves 73.7 in macro F1 score. 
We then tested performance by repositioning the wrist IMU at other locations along the arm. 
As shown in \figref{fig:study1-activity}, relocation of the IMU to the hand or lower arm (below the elbow) has minimal impact on overall performance, while relocation to the upper arm decreases activity recognition accuracy.
This decrease is reasonable, as fine-grained hand or lower arm motions cannot be directly captured from an upper arm placement, but must instead be inferred by the model, naturally creating additional challenges for activity recognition.

\subsection{Evaluation~\#2: Motion Sensing at Typical Locations Using Existing Datasets}
\subsubsection{Body Pose Tracking from 6 IMUs}

We systematically analyzed the performance of \proposed{} in tracking body poses with state-of-the-art models using the TotalCapture dataset ~\cite{trumble2017total} for ease of comparison with previous work.
Following Wouwe~\etal~\cite{DBLP:conf/cvpr/WouweLFDL24}, we evaluated using all six real IMU sensors placed in the pelvis, head, wrists, and shanks, and also subsets of them. 
In this section, we use different terminology for sensor locations to remain consistent with the datasets used.
In consistency with previous work~\cite{DBLP:journals/tog/YiZ021, DBLP:conf/cvpr/WouweLFDL24, DBLP:conf/siggraph/Yi0X24, DBLP:conf/cvpr/YiZHSGT022}, we used TotalCapture only for testing and excluded it from our training data set at all stages.

\begin{table}[t]
\centering
\caption{Comparison of pose estimation approaches using the conventional 6 IMUs (pelvis, head, left wrist, right wrist, left shank, right shank) on TotalCapture dataset. Lower Global Angular Error (GAE) is better.}
\Description{This table summarizes the different pose estimation approaches, including their compatible placement and their Global Angular Error using 6 IMUs.}
\begin{tabular}{lcc}
\toprule
\textbf{Method} & \textbf{Compatible Placement} & \textbf{GAE (deg)} \\
\midrule
Transpose~\cite{DBLP:journals/tog/YiZ021} & Fixed 6 & 16.1 \\
PIP~\cite{DBLP:conf/cvpr/YiZHSGT022} & Fixed 6 & 14.4 \\
PNP~\cite{DBLP:conf/siggraph/Yi0X24} & Fixed 6 & 10.4 \\
DiffusionPoser~\cite{DBLP:conf/cvpr/WouweLFDL24} & Select From 13 & 14.4 \\
\textbf{IMUCoCo} & \textbf{Anywhere on Body} & \textbf{14.0} \\
\bottomrule
\label{table:result6imu}
\end{tabular}
\end{table}
The comparison using the conventional 6 IMUs (pelvis, head, left wrist, right wrist, left shank, right shank) on the TotalCapture datasets is summarized in~\tabref{table:result6imu}. 
Overall, IMUCoCo with DTP still achieves reasonable performance, even when compared with pose estimation models specifically designed for 6-IMU setups. 
We believe that the 6-IMU configuration offers a unique advantage for estimating pose, as it converts all the leaf IMU to be relative to the pelvis IMU, which eliminates the signal variation caused by different facing directions. This normalization approach is used in several prior works, such as PNP~\cite{DBLP:conf/siggraph/Yi0X24}, DynaIP~\cite{zhang2024dynamic}, PIP~\cite{DBLP:conf/cvpr/YiZHSGT022}, and Transpose~\cite{DBLP:journals/tog/YiZ021}.
For Diffusion Poser~\cite{DBLP:conf/cvpr/WouweLFDL24}, IMUPoser~\cite{DBLP:conf/chi/MollynAG0A23}, MobilePoser~\cite{DBLP:conf/uist/XuGHA24}, and our work, we can only use the global frame IMU measurements as these use cases do not always have a pelvis IMU. 

\subsubsection{Body Pose Tracking from Flexible IMUs}

\begin{table}[t]
\caption{Comparison of Global Angular Error (GAE) between IMUCoCo + DTP and DiffusionPoser under different test placements using the TotalCapture dataset. Lower is better.
\small{(P = pelvis; H = head; RLA = right lower arm; LLA = left lower arm; RLL = right lower leg; LLL = left lower leg)}}
\Description{This table summarizes the Global Angular Error for different IMU placements on the TotalCapture dataset.}
\centering
\begin{tabular}{p{3.52cm}cc}
\toprule
\textbf{Test Placement} & \textbf{IMUCoCo} & \textbf{DiffusionPoser~\cite{DBLP:conf/cvpr/WouweLFDL24}} \\
\midrule
All 6 IMUs & \textbf{14.0} & 14.4 \\
P+H+RLA+LLA+RLL & \textbf{15.7} & 19.4 \\
P+H+RLL+LLL & \textbf{21.8} & 24.9 \\
P+RLL+LLL & \textbf{23.8} & 36.4 \\
RLL+LLL & \textbf{26.6} & 39.2 \\
H & \textbf{32.0} & 39.2 \\
\bottomrule
\end{tabular}
\label{tbl:gae_comparison}
\end{table}

Next, we examined the performance of ~\proposed{} using a flexible combination from a set of sensor locations. DiffusionPoser~\cite{DBLP:conf/cvpr/WouweLFDL24} employs a generative diffusion process to infer the missing sensor and pose features and supports IMU placement from at most 13 possible locations. To the best of our knowledge, DiffusionPoser is currently the most flexible IMU-based full-body pose method. 
\tabref{tbl:gae_comparison} summarizes the comparison result using 6, 5, 3, 2, and 1 IMU sensors on the TotalCapture dataset.
\proposed{} performs better on all the listed sensor combinations than DiffusionPoser~\cite{DBLP:conf/cvpr/WouweLFDL24}. 

\subsection{Ablation Analysis}
In this section, we provide a more detailed analysis of the individual components of the IMUCoCo system and its application performance. 
The evaluation is conducted on both the benchmark dataset and our custom datasets. 
We first implemented DTP as our pose estimation model using the same configuration in the experiment (\secref{subsection:downstream_task}), but without using IMUCoCo at all, denoted as \textbf{w/o IMUCoCo}.
To ensure the model has the same dimension and depth, we mapped the original IMU inputs using 24 different linear projections so that it shared the same dimensions as the extracted features from IMUCoCo.
We hypothesize that while these models should still provide reasonable performance on the benchmark datasets with the same 6-IMU configurations, they will generalize poorly to new sensor locations in our custom dataset.

Then, we specifically tested the detailed designs of the IMUCoCo model by evaluating the impact of the sensor coordinate information. 
For this testing, we drop the sensor coordinate information by always setting it to zero; we name this approach IMUCoCo w/o Sensor Coordinates or \textbf{w/o SC}. 
In this test, each joint node will transfer the matched IMU signal, but without knowing where exactly the sensor comes from.

Regardless of the models used in the ablation analysis, the input dimensions were still allocated the same using the \proposed{}'s transfer loss map based on the spatial location of the provided real IMU sensors, as otherwise, the downstream model will have no information on where to take the input of various IMU sensors to produce a meaningful comparison. 
\begin{table}[t]
\caption{Ablation analysis of IMUCoCo for pose estimation measured in Global Angular Error (GAE) on TotalCapture datasets. Lower is better.
\small{(P = pelvis; H = head; RLA = right lower arm; LLA = left lower arm; RLL = right lower leg; LLL = left lower leg)}
}
\Description{This table shows the results of the ablation analysis of IMUCoCo on the TotalCapture dataset}
\label{tab:ablation_tc}
\centering
\begin{tabular}{p{2.6cm}ccc}
\toprule
\textbf{Test Placement} & \textbf{IMUCoCo} & \textbf{w/o IMUCoCo} & \textbf{w/o SC} \\
\midrule
All 6 IMUs & 14.0 & 16.6 & 26.8 \\
P+H+RLA+LLA+RLL     & 15.7 & 18.1 & 26.5 \\
P+H+RLL+LLL         & 21.9 & 33.8 & 27.2 \\
P+RLL+LLL           & 23.8 & 35.7 & 27.9 \\
RLL+LLL             & 26.7 & 42.9 & 35.0 \\
H                   & 32.0 & 43.5 & 37.2 \\
\bottomrule
\end{tabular}
\end{table}

\tabref{tab:ablation_tc} shows the GAE on the approaches above for pose estimation on the TotalCapture dataset. 
In comparison with DTP w/o IMUCoCo, which is specifically trained only using the 6-IMU configuration as used for this dataset, \proposed{} still achieves better performance. 
We attribute this improvement to the known IMUs, primarily to the architecture of \proposed{}, which not only enables flexibility in testing, but its scalable architecture during the pre-training phase allows utilizing the information from rich augmented mesh IMU signals, which helps to learn a more robust motion feature representation. 
Comparing w/o SC to w/o IMUCoCo, we found a significant degradation of performance if using ~\proposed{} without providing the correct IMU device coordinate.
In this case, the joint node module in ~\proposed{} confuses how to transfer the provided signal correctly. This also verifies that \proposed{} has learned to adapt based on the sensor coordinate of IMU devices. 

\begin{table}[t]
\centering
\caption{Ablation analysis of IMUCoCo for pose estimation measured in Global Angular Error (GAE) on our custom datasets. Lower is better. The arrow indicates moving one standard placement (wrist, thigh, and ear) to another placement.}
\Description{This table shows the results of the ablation analysis of IMUCoCo on the custom dataset.}
\begin{tabular}{p{2.6cm}ccc}
\toprule
\textbf{Test Placement} & \textbf{IMUCoCo} & \textbf{w/o IMUCoCo} & \textbf{w/o SC} \\
\midrule
wrist, pocket, ear & 27.6 & 43.4 & 35.0 \\
wrist $\rightarrow$ arm & 28.6 & 43.5 & 34.9 \\
pocket $\rightarrow$ leg & 29.3 & 42.0 & 36.2 \\
ear $\rightarrow$ torso & 29.1 & 42.9 & 35.2 \\
\bottomrule
\end{tabular}
\label{tab:ablation_custom}
\end{table}
\tabref{tab:ablation_custom} presents GAE for different IMU placement configurations on our custom dataset, where each row corresponds to a variant of the standard 3-IMU setting (wrist, thigh pocket, ear) with one IMU repositioned at each time. 
The model trained without IMUCoCo fails immediately across all configurations, as it cannot generalize to these unseen or altered IMU placements from its training. Note that this model is trained using the 6-IMU combinations, rather than the wrist, thigh pocket, and ear.
The version without sensor coordinate (w/o SC) still shows significant degradation in comparison with \proposed{}, confirming that \proposed{} utilizes the sensor coordinate information to adapt accurately to the input IMU signals.

\section{Application Scenarios}

\proposed{} allows users to put their devices with IMUs anywhere they prefer. 
This capability opens up a wide variety of novel applications (\eg~\figref{fig:teaser}).

\proposed{} allows end-users to utilize the myriad of smart devices they may have that are equipped with an IMU. Prior work has focused on enabling motion and pose sensing from the most common devices. We support the user to use the device they have or need. 
For example, runners typically do not like to keep their phones in their pants pockets.
We enable them to track their running form, if they like, from an armband placement. 
Similarly, if a user has a necklace or smart innerwear, which sometimes has an IMU built into the waistband, we do not need to retrain a model that is tuned to estimate torso movements from the ear.

\proposed{} adapts to shifts in sensor placement that may occur throughout the day.
For example, a user might prefer to keep their phone in different pockets throughout the day or in various activities. As this user moves their phone from their pant pocket to their sweatshirt's kangaroo pocket to their pants' back pocket to the forearm pocket in their ski jacket, \proposed{} adapts to these changing placements and provides accurate pose and activity estimates for those specific activities. 
As long as \proposed{} is aware of the active sensor's location, it can adjust to these changes without the need for switching models. 
This flexibility in changing device locations dynamically supports adaptive sensing, aligning with the variable contexts of daily activities.

Moreover, \proposed{} also supports changing or suggesting placements to meet the user's specific needs. 
Khurana~\etal~\cite{khurana2019detachable}proposed a detachable smartwatch that can be placed on different parts of the body depending on the end need.
\proposed{} can now support such scenarios with accurate motion sensing. 
For example, a person with two IMU devices can record their footwork when playing soccer by attaching them to two thigh pockets (See \figref{fig:teaser}C).
If the user's focus then shifts to upper-body posture, the user can move the sensor to their chest or arms.
Or, if the user wants to track their squat form more accurately, we can recommend an optimal placement for the watch (perhaps wear it as an anklet) and help the user ensure that their knees do not go beyond their toes. 
Throughout these adjustments, \proposed{} consistently provides tracking and analysis without the need to switch or retrain models.
No doubt, the placement recommendations will need to take into account the user's device form factor and available adapters/straps. 

\section{Limitations and Future Work}
\label{sec: limitation}
The current implementation of \proposed{} is not without limitations.
First, the current synthetic IMU data generation method is based on the assumption of a rigid human body.
However, in practical scenarios, IMU devices are often attached to clothing, which introduces variability due to the flexibility and movement of the fabric.
This discrepancy between the synthetic data and real-world data results in inaccuracies in the signals, as the synthetic model does not account for the flux caused by clothing.
Future work should continue refining the synthesis process to include models that simulate the impact of clothing ~\cite{zuo2024loose} and other factors such as contacting objects.
Enhancing the realism of synthetic datasets will improve the robustness and applicability of the system across more realistic human activity scenarios.

Secondly, it is challenging to accurately estimate full-body movements from sparse sensors placed at any location, especially when the point of interest is far from the sensor.
When the placement is not ideal for an application, \proposed{} will still produce its estimate but with higher errors.
The model outputs can involve hallucinations, relying on learned correlations within the training data rather than on direct observations.
This approach may suffice for certain applications, but it falls short when precise and reliable full-body tracking is required.
Future improvements should aim to increase the transparency of model predictions by visualizing which body parts \proposed{} estimates with confidence and which parts are less accurately represented.
This enhancement will provide users with a clearer understanding of the model's capabilities and limitations, facilitating a more informed application of the technology in complex scenarios. 
Moreover, it is crucial to understand what potential placement might be preferable by users and by applications. Such consideration extends beyond optimizing performance to include factors such as comfort and the stability of device attachment. 

Finally, future work shall consider further simplifying the specification of the on-body IMU placement by exploring an intuitive interface or calibration step.
For instance, Sztyler and Stuckenschmidt~\cite{DBLP:conf/percom/SztylerS16} have proposed a method to recognize the on-body location of each device from IMU signals.
In addition, using a dedicated network or sensing technology to identify the location of the device might also be possible. 
These developments would lower the barriers for configuring \proposed{} and support the application scenarios we discussed above.

\section{Conclusion}

The integration of IMUs into consumer devices has significantly advanced the field of human motion and pose estimation.
Our evaluations demonstrated that \proposed{} can effectively utilize IMUs placed at flexible locations on the body by projecting their signals onto a placement-adaptive representation, which can be adjusted for downstream tasks such as pose tracking and activity recognition.
\proposed{} offers numerous advantages that enhance the practicality of on-body motion sensing: it accommodates devices at unconventional locations, allows for the seamless movement of devices throughout the day, and optimizes sensor placement based on specific activities, all while using a single model without the need for retraining.
We believe our approach enables a wide range of practical applications, from personal fitness tracking to rehabilitation.
More broadly, we hope our work encourages further research in scaling up the development of IMU-based sensing models for human motion understanding.

\begin{acks}
This work was supported in part by the Center for Machine Learning and Health at CMU, Samsung, and Masason Foundation. 
We also thank Vasco Xu, Karan Ahuja, and Vimal Mollyn for their valuable advice and for sharing their implementations, and David Lindlbauer and Yi Fei Cheng for their assistance with motion capture equipment.
\end{acks}

\bibliographystyle{ACM-Reference-Format}
\bibliography{paper}
\balance 
\newpage
\appendix

\section{System Details}
\label{app:system}
\subsection{Virtual IMU Synthesis}
\label{app:system-virtual}
For virtual mesh IMU, we synthesize the orientation based on faces. 
Specifically, for each vertex on the human body mesh, we took the face norm as all the faces connecting to this vertex and took the average norm vector of the faces as the $y$-axis.
The $y$-axis then always points outside the body surface and is perpendicular to the tangent plane of the body surface.
$z$-axis is then determined by the direction orthogonal to the plane formed by the bone direction and the face norm, and the $x$-axis is orthogonal to both $y$ and $z$-axis. 

We also synthesize IMU from the joints in addition to the vertices on the body surface to create a reference for \proposed{} during training.
To synthesize IMU and kinematics attributes for joints, we did not use the original joint positions.
This is because the original joint's acceleration and orientation do not fall into the same distribution as an IMU attached to the surface.
For example, the orientation at the right elbow joint is similar to the orientation of an IMU attached to the lower arm, but its acceleration is more similar to the acceleration of an IMU attached to the upper arm. 
Thus, we modified the locations to sample the acceleration, velocity, and position of the joint from its child joint while keeping the joint orientation from its own.
For example, the joint node at the right elbow will synthesize its acceleration from the right wrist, while keeping its own orientation at the right elbow. 
The coordinates of this joint node are defined by the root-centered position of the right wrist instead of the right elbow. 
For the leaf joint in the head, hands, and feet, we created five additional vertices on the body on the top of the head, fingertips, and toes to calculate acceleration, velocity, and position.
\begin{figure}
    \centering
    \includegraphics[width=0.7\linewidth]{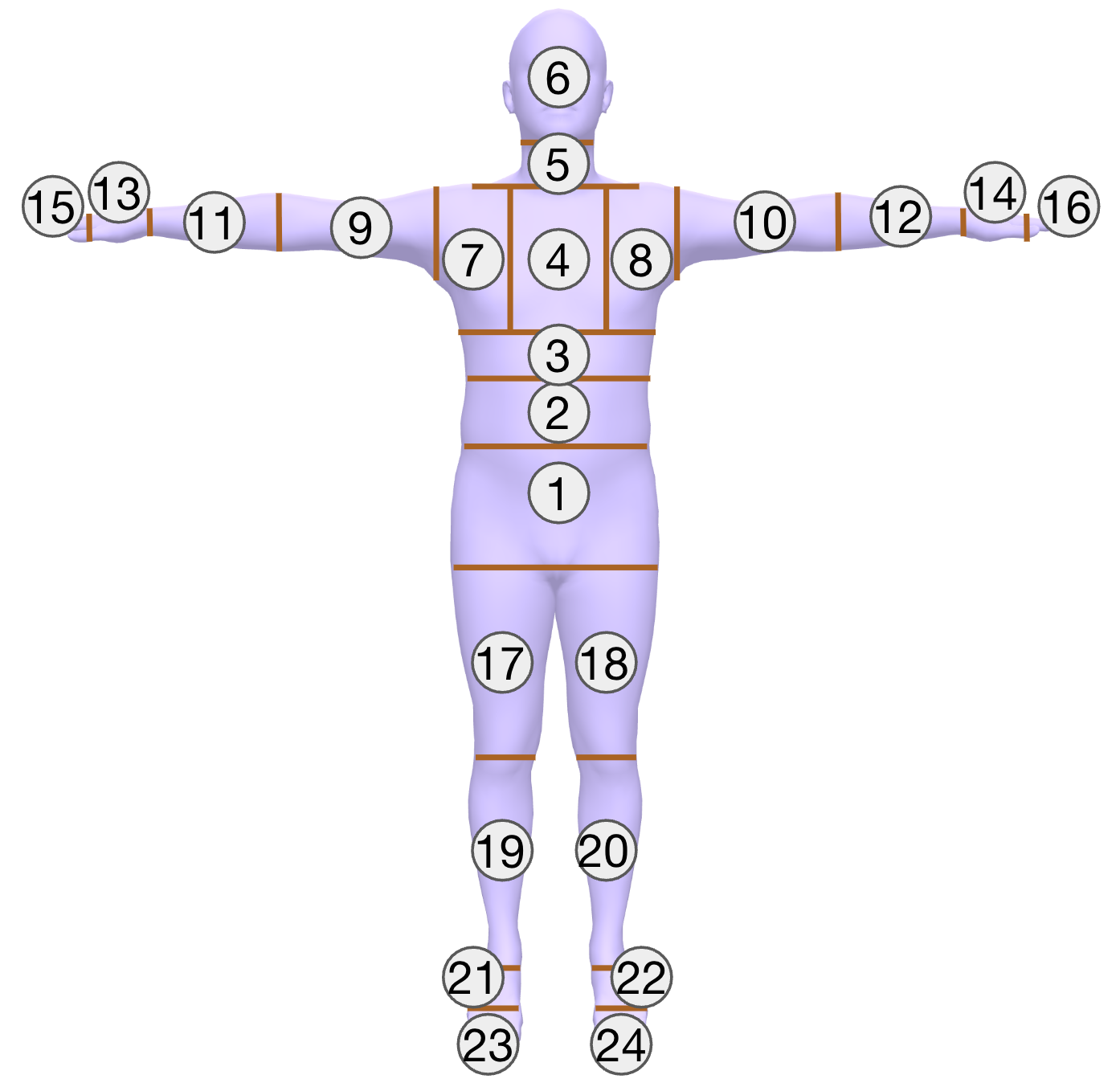}
    \caption{Illustration of categorizing body regions based on horizontal and vertical planes passing through the joints at T pose.}
    \Description{Illustration of body region categorization using joint-based segmentation in a T-pose.}
    \label{fig:region_category}
\end{figure}
\begin{figure*}
    \centering
    \includegraphics[width=0.93\linewidth]{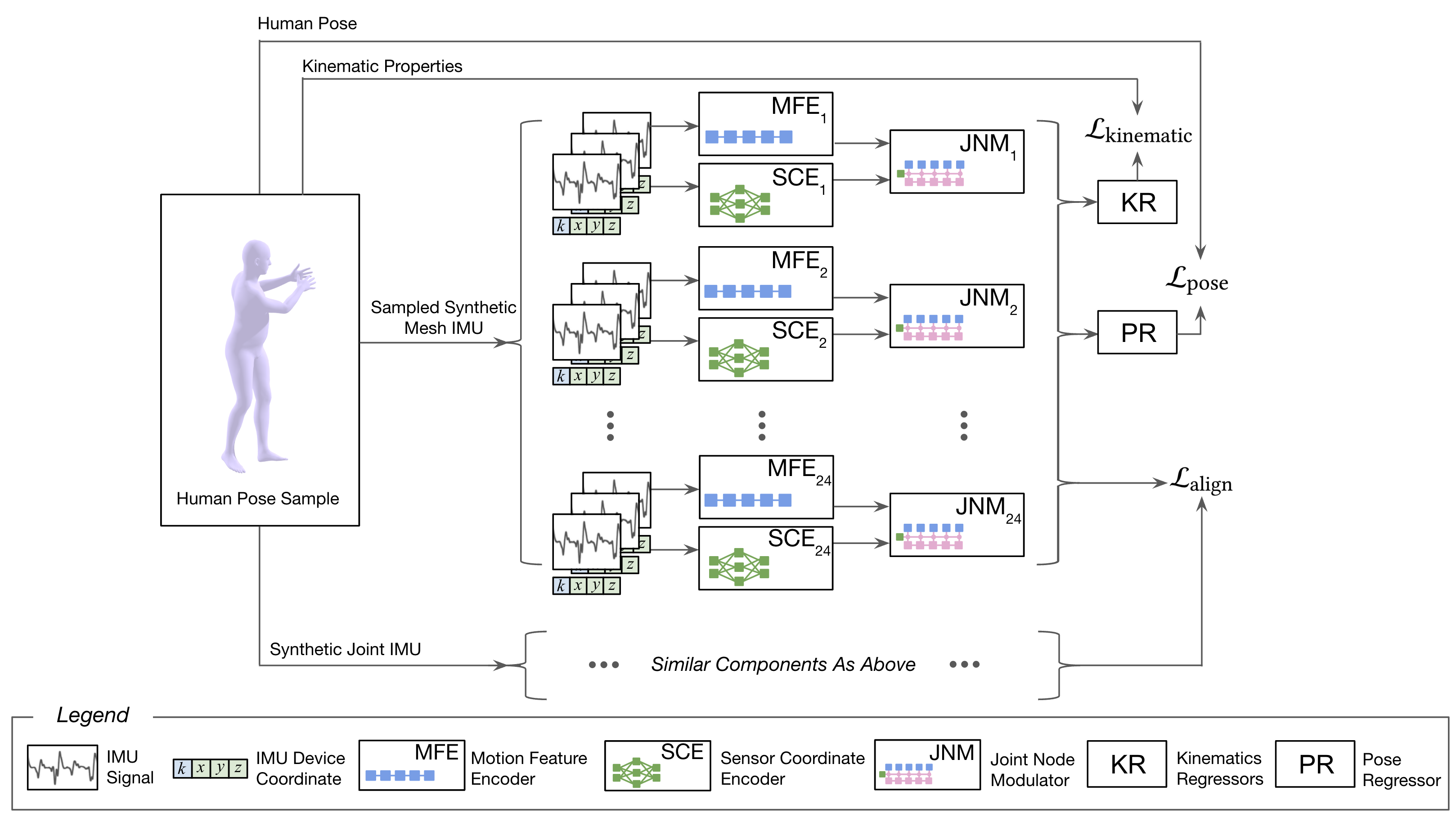}
    \caption{Overview of the \proposed{} system during training time.}
    \Description{The figure presents the full training pipeline of IMUCoCo, illustrating how synthetic IMU signals are generated from human pose samples and how those signals are used to learn robust, placement-adaptive representations.}
    \label{fig:training_system}
\end{figure*}
\subsection{Training Architecture and Procedure} \label{appendix:training}
\figref{fig:region_category} shows the categorization $k$ by sensor coordinates. 
The regions are partitioned by horizontal and vertical planes passing through the joints at T-pose. 
Vertices in each of the categorized regions often exhibit more similarities than those in distant region categories. 
Together with frequency-based positional encoding of the coordinates, the concatenated features are fed into the MLP layers to derive placement codes.   

\figref{fig:training_system} shows the full architecture, including the auxiliary components and mesh IMU sampling, during the training process. 

We used a two-phase training procedure. In the first phase, we warm up the model using joint IMUs. The advantage of this phase lies in the small size of the input (24 joint signals), but at the same time, this input contains enough high-quality information that can lead to an accurate description of the full-body motion.
Therefore, this step allows the model to provide a warm start to all the modules, especially to learn accurate KRs and PR that are used for supervision in the next phase. 

In phase two, we extensively sample the full-body dense mesh virtual IMUs from the body surface. 
We freeze the KRs, as each sample is trained repeatedly on the 24 joint nodes; otherwise, the KRs will overfit to this batch quickly. 
To focus on the regions that are physically plausible for transference, we apply a weighted stratified sampling scheme.
We calculate a weight decay based on the number of hops from the sampled point to each joint node, as well as the initial vertex density of the articulated pose model, so that fewer hops lead to higher chances of being sampled.
For each motion sequence, we first perform a forward pass using the 24 virtual joint IMUs, and save this representation as a buffer. Next, for each joint node, we sample a large amount (384 points per motion sequence in our training) of mesh virtual IMUs for each joint node, and perform a forward pass. 
The inferred feature from this forward pass will replace the corresponding joint feature from the buffer, and together, pass through the full-body pose regression.
The advantage of this buffered approach is that gradient updates are applied to each joint node for each forward pass using the mesh virtual IMU, without the need to save the gradient and wait until all 24 joint nodes finish their forward pass (for which our GPU does not have enough memory to execute).
We then applied the kinematic loss to both the output using the joint virtual IMU and the mesh virtual IMU, as well as the alignment loss between the two. 

\end{document}